# Empirical exploration of air traffic and human dynamics in terminal airspaces

Lei Yang [a], Suwan Yin [b,a,*], Minghua Hu [a], Ke Han [b,a], Honghai Zhang [a]

[a] College of Civil Aviation, Nanjing University of Aeronautics and Astronautics, Nanjing 210016, China
[b] Department of Civil and Environmental Engineering, Imperial College London, SW7 2BU, UK

**Abstract:** Air traffic is widely known as a complex, task-critical techno-social system, with numerous interactions between airspace, procedures, aircraft and air traffic controllers. In order to develop and deploy high-level operational concepts and automation systems scientifically and effectively, it is essential to conduct an in-depth investigation on the intrinsic traffic-human dynamics and characteristics, which is not widely seen in the literature. To fill this gap, we propose a multi-layer network to model and analyze air traffic systems. A *Route-based Airspace Network* (RAN) and *Flight Trajectory Network* (FTN) encapsulate critical physical and operational characteristics; an *Integrated Flow-Driven Network* (IFDN) and *Interrelated Conflict-Communication Network* (ICCN) are formulated to represent air traffic flow transmissions and intervention from air traffic controllers, respectively. Furthermore, a set of analytical metrics including network variables, complex network attributes, controllers' cognitive complexity, and chaotic metrics are introduced and applied in a case study of Guangzhou terminal airspace. Empirical results show the existence of fundamental diagram and macroscopic fundamental diagram at the route, sector and terminal levels. Moreover, the dynamics and underlying mechanisms of "ATCOs-flow" interactions are revealed and interpreted by adaptive meta-cognition strategies based on network analysis of the ICCN. Finally, at the system level, chaos is identified in conflict system and human behavioral system when traffic switch to the semi-stable or congested phase. This study offers analytical tools for understanding the complex human-flow interactions at potentially a broad range of air traffic systems, and underpins future developments and automation of intelligent air traffic management systems.

**Keywords:** Terminal airspace, air traffic flow, air traffic controller, nonlinear dynamics, chaos

## Nomenclature

| | | | |
|---|---|---|---|
| AFR: | *Average Flow Rate* | ICCN: | *Interrelated Conflict-Communication Network* |
| ASBU: | *Aviation System Block Upgrades* | ICN: | *Interventional Communication Network* |
| ATCO: | *Air Traffic Controller* | IFDN: | *Integrated Flow-Driven Network* |
| ATV: | *Average Traffic Volume* | NextGen: | *Next Generation Air Transportation System* |
| CL: | *Communication Load* | OFP: | *Operational Flight Path* |
| CSN: | *Conflict Situation Network* | RAN: | *Route-Based Airspace Network* |
| EAD: | *Equivalent Average Density* | SESAR: | *Single European Sky ATM Research* |
| EAS: | *Equivalent Average Speed* | SFP: | *Standard Flight Path* |
| FTN: | *Flight Trajectory Network* | SSPC: | *Solution Space-based Perceived Complexity* |

## 1. Introduction

Worldwide ATM system is undergoing the process of upgrading and transformation to cope with increasing air traffic demand and congestion especially in high-density airports and surrounding airspaces. Within the strategic planning of ATM systems like SESAR, NextGen and ASBU, numerous advanced operational concepts, e.g. "ATM Network Management", "User Driven Prioritization Process", "Flow Contingency Management" and "Complexity Management", are proposed to enhance the system-wide performance and reduce the propagation of congestion. Air traffic systems are typical examples of complex techno-social systems, the management of which requires a careful understanding of the subtle interplay between technological infrastructure and human behavior (Monechi et al., 2015). In order to develop and deploy high-level operational concepts and automation systems in an effective way, it is essential to conduct an in-depth investigation of the intrinsic air traffic dynamics, by revealing temporo-spatial air traffic characteristics and uncovering the intrinsic "human" (ATCOs) and "flow" (aircraft) interactions.

Flow congestion is one of the key manifestations of air traffic dynamics (Hu et al., 2013). Nevertheless, studies on the generation, accumulation, propagation, and dissipation of air traffic congestion are not widely reported. Delay has been studied as a key performance indicator of airspace capacity since the 1940s (Bowen et al., 1948, David et al., 1998). The classic exponential relationship between capacity utilization and delay presents the most fundamental relationship between demand and supply, and has guided ATM research for decades that followed (FAA, 1983). Some recent findings suggest refined demand-supply relationship at airports based on empirical data (Ezaki et al, 2014, Simaiakis et al., 2014), which underpins novel control strategies for congestion mitigation. In order to capture the spatial effect of congestion, delay propagation models receive much attention. Single flight delay is modelled using time-based petri network to analyze the chain reaction (Li et al., 2008), and provides the basis for flight delay prediction and alerting (Xu et al., 2009). Furthermore, the mechanisms for the emergence and accumulation of delays are studied based on a chain reaction of multi-flight delay model (Hua et al. 2007) and analytical econometric approach (Kafle et al, 2016) in airport networks.

*Corresponding author
*Email address*: laneyoung49@hotmail.com (L. Yang), hyy_170304@163.com (S. Yin), minghuahu@263.net (M. Hu), k.han@imperial.ac.uk (K. Han), zhh0913@163.com (H. Zhang)

The flow dynamics of vehicle traffic have motivated many air traffic flow models in recent years, in order to predict the aggregate effect on air traffic delays and to support large-scale flow management. Bayen et al. (2006) introduce partial differential equations (Lighthill et al., 1955) to the prediction of air traffic flow propagation, and the control of air traffic flow along one-dimensional air-routes in National Airspace Networks. Inspired by the cell transmission model (CTM) of vehicular traffic (Daganzo, 1994), 1D (Menon et al., 2004) and 2D (Menon et al., 2006) cell models are derived by discretizing the partial differential equations. Building on the models of Menon et al., Large-capacity CTM is proposed by Sun et al. (2008) and Wei et al. (2013) to model large-scale air traffic networks by distinguishing link and cell levels for each flight path. Cao et al. (2011) develop a Link Transmission Model without discretization on the cell level to improve computational efficiency. Zhang et al. (2014) propose a CTM-based flow model for terminal airspace, via a hypothesized "flow-density-velocity" relationship (or the fundamental diagram, see Kerner, 2009) of terminal air traffic. The primary focus of the abovementioned studies is the control of air traffic based on airspace or route capacity, with little attention given to the empirical validity of the hypothesized flow-density-velocity relationships, which play a vital role in the aggregate behavior of air traffic and are essential to the effectiveness of such controls.

With the increased workload of ATCOs, significant efforts have been devoted to the modeling and analysis of human dynamics. Odoni et al. (1997) categorize ATCOs' behavioral models into macroscopic and microscopic classes. Macroscopic models are high-level analytical models that analyze individual ATCO's performance by establishing input-output transfer mechanisms, treating the internal information process as a black box. Examples of this type include workload model (Farmer et al., 2003), crossover model for manual control tasks (McRuer et al., 1967) and human response delay model (Ren et al., 2005). On the other hand, microscopic models attempt to describe the cognitive processes (e.g. attention resource assignment, memory usage, situation awareness, decision making and monitoring) in greater detail; examples include CT-ATC (Kallus et al., 1999), MoFL (Eyferth et al., 2003), and Apex (Lee et al., 2005). The quantitative modeling of ATCOs' behavioral dynamics sheds light on the fundamental rules that human follow when executing tasks, and provides tools to refine traditional flow-based air traffic modeling. Wang et al. (2013) propose an empirical method to study ATCOs' high-level dynamics by analyzing their communication intervals, and suggest that the distribution of communication intervals follows the power-law distribution. A follow-up study (Wang et al., 2016) introduces a network approach to the study of ATCOs' communication data, and further reveals unique patterns in the ATCOs' activities. However, these studies rely exclusively on the controllers' communication data, ignoring their coupling and interdependencies with the air traffic data; thus no understandings of the ATCO-flow evolutionary dynamics are extracted.

There exist a number of studies that attempt to bridge the gap between traffic states and ATCOs' cognition complexity, by using complexity methods that interpret the "human-flow" dynamics. To this end, a series of traffic flow complexity metrics are developed in the last two decades. Classic metrics include Static Density (Sridhar et al., 1998), Dynamic Density (Laudeman et al., 1998), Tactical Load Smoother (Whiteley, 1999), Input-Output (Lee et al., 2009), Lyapunov exponent of trajectory dynamics (Puechmorel et al., 2009), and Solution space-based metrics (d'Engelbronner et al., 2015). By finding the best match between traffic flow complexity measurements and ATCOs' workload, the weights of relevant indicators (e.g. traffic density, aircraft type mixture, potential conflict) are calibrated. The weight can be treated as the impact of each traffic scenario on the ATCO's cognition. Monechi et al. (2015) model a large-scale air transportation system as a complex, dynamic network of flights controlled by humans, and analyze the probability distributions of delay times and potential conflicts. As the development of 4D trajectory operation, Zohrevandi et al. (2016) designed numbers of scenarios to measure how the level of 4D equipped affect controllers' task load in terminal and en route sectors. Corver et al. (2016) studies how trajectory uncertainties impact controllers' workload under various traffic density and conflict situations. This line of research, despite the fresh perspective on the relationship between traffic flow and cognitive measures, are inadequate to analyze the evolutionary dynamics of "human-flow" and the mechanisms through which they interact.

As a summary of the literature review above, existing studies focus on the modeling and evaluation of air traffic systems without much effort to understand their inherent ATCO-flow dynamics and rules that govern their co-evolution. There is a lack of quantitative characterization of the complexity of the air traffic system from the perspective of human and aircraft interactions. In order to characterize and interpret the coupling "ATCOs-flow" dynamics and their interaction and co-evolution at both macroscopic and microscopic levels, we propose a novel multi-layer network approach. The multi-layer network consists of two physical networks: *Route-based Airspace Network* (RAN) and *Flight Trajectory Network* (FTN), and two functional networks: *Integrated Flow-Driven Network* (IFDN) and *Interrelated Conflict-Communication Network* (ICCN). In order to underpin such a multi-layer network framework, and to reveal internal "ATCOs-flow" dynamics, we develop three classes of analytical metrics: (i) network flow metrics for IFDN; (ii) complexity metrics for ICCN; and (3) chaos metrics for system-level characterization. The main contribution of this paper includes not only the multi-layer network model, but also the following main findings through an empirical study of a terminal airspace in China. The content and logical flow of this paper are shown in Figure 1.

(1) **[Empirical FD and MFD]**. At the route segment level based on IFDN, we use empirical data, for the first time, to establish the flow-density-speed Fundamental Diagram (FD) (Daganzo, 1994, 1995). Four distinct flow phases are identified: free-flow, smooth, semi-stable, and congested. In addition, at the network level we use empirical data to reveal the Macroscopic Fundamental Diagram (MFD), which captures the "demand-supply" relationships and "arrival-departure" interactions. To the best of our knowledge, this is the first reported MFD instance for terminal airspaces.

(2) **[ATCOs-flow interactions]**. At the sector level, by analyzing the degree distribution and motifs[1] of ICCN, meta-cognition strategies, *pre-activated*, *inhibition* and *stress* during the transitions of phases mentioned in (1), are shown to be the underlying mechanisms of ATCOs-flow co-evolution.

(3) **[Chaotic dynamics at the system level]**. Chaos is found in both air traffic flow dynamics and ATCOs' activities by analyzing potential traffic conflict and communication activities. Moreover, we show that chaos emerges at unstable phases, i.e. semi-stable and congested phases.

The linkage between FD/MFD and the ATCO-flow interactions can be illustrated as follows. Firstly, MFDs at the sector level, which are used to analyze human-flow interaction (Section 4.3), are inherited from MFD at the terminal level (see Section 4.2.2). Secondly, the description of FD and discussions on phase transitions presented in Section 4.2.1 provide reference for phase identification of integrated "ATCO-flow" performance at the sector level. In other words, FD and MFD underpin the study of the "ATCO-flow" interactions.

---

[1] Network motifs, which is defined as the patterns of interconnections occurring in complex networks at numbers that are significantly higher than those in randomized networks, can be also simply regarded as the building blocks of large complex networks (Milo et al., 2002).

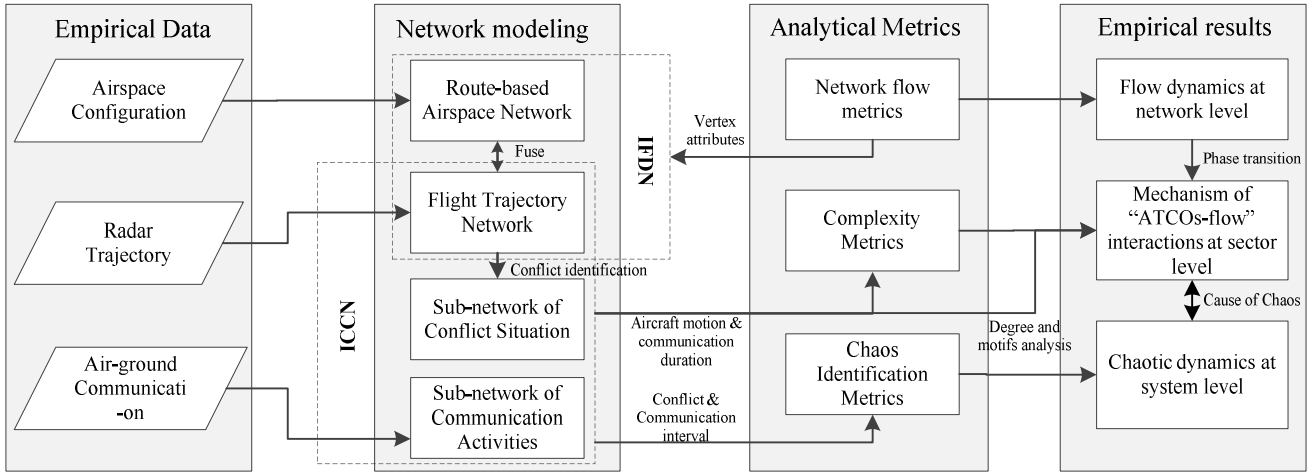

Figure 1. Framework of proposed network-based air traffic dynamics analysis.

This paper proposes an integrated framework for analyzing ATCO-flow dynamics, and makes significant empirical findings through a case study of a high-density terminal airspace in China. With regard to future air traffic management, in which customized and intelligent automations will play a vital role to support ATCOs and managers with decision-making, this study has profound impacts as follows.

(a) It underpins advanced modeling and management. The calibrated FDs can be utilized to support mesoscopic dynamic modeling of terminal air traffic, such as those studies mentioned in the literature review. Meanwhile, the MFDs can be integrated with large-scaled flow management by treating the terminal airspace as a high-capacity cell.

(b) It supports automation design. The underlying mechanisms of "ATCOs-flow" interactions revealed by this paper provide improved understanding of the ATCOs' adaptive behavioral patterns, and serve as the foundation for customized and intelligent automations with higher acceptance rate by ATCOs.

(c) It enhances system predictability. The potential chaotic nature of the air traffic system, as revealed in this paper, renders its prediction a great challenge due to the randomness. Chaotic prediction methods such as the largest Lyapunov exponent can be used during high-demand scenarios to enhance traffic management by forecasting potential conflicts and ATCOs' workload.

The rest of this paper is organized as follows. Section 2 presents the multi-layer network model for human-flow dynamics. In Section 3, a number of metrics are introduced for the empirical evaluation of air traffic dynamics. Section 4 carries out a case study of Guangzhou terminal airspace to investigate the non-linear and complex dynamics of air traffic. Finally, some conclusions are provided in Section 5.

## 2. A multi-layer network model for integrated "human-flow" analysis

In sector-based operations, complex air traffic flow dynamics arise from non-linear interactions among sector configuration, air traffic, and ATCOs. In order to conduct integrated analyses on the "human-flow" dynamic, we develop a novel, analytical framework based on multi-layer networks. The framework consists of two physical networks: *Route-base Airspace Network (RAN), Flight Trajectory Network (FTN)*, and two functional networks: *Integrated Flow-Driven Network (IFDN)* and *Interrelated Conflict-Communication Network (ICCN)*. These networks are illustrated in Figure 2 and elaborated in the sections below.

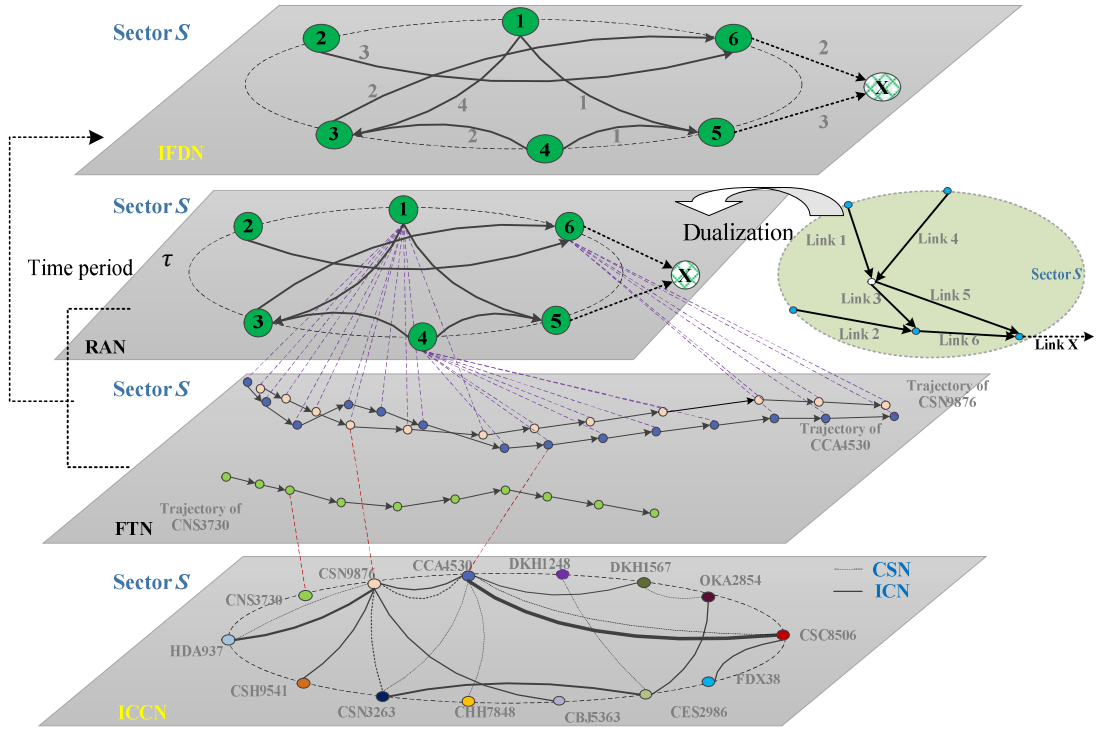

Figure 2. Multi-layer network for "human-flow" analysis.

## 2.1. Route-based airspace network (RAN)

Conventionally, airspace networks are modeled as graphs with waypoints as vertices and route segments (links) as edges. In this section, however, we construct the dual network of the conventional network, and call it the RAN and denote it as $G_{RAN}(V_{RAN}, E_{RAN})$. The set of paths in the RAN is denoted $\mathcal{P}_{RAN}$. As illustrated in Figure 3, the vertices in the dual network represent the links in the original network, and the directed connectivity in the dual network reflects the adjacency of the links in the original network. For example, vertex 8 is connected to 9 in the dual network means that edge 8 is immediately upstream of edge 9 in the original network. Some benefits of invoking the dual network are as follows. Firstly, the RAN-FTN mapping can be expressed in terms of inter-network edges for ease of demonstration, as shown in Section 2.3. Secondly, for consistency, IFDN and RAN have the same set of vertices if the dual network is employed for RAN. Finally, compared to the conventional original network, the dual network of RAN provides a structure for IFDN with a better capability of modelling the flow transmission between links for all kinds of route configurations (e.g., merge, diverge or crossover).

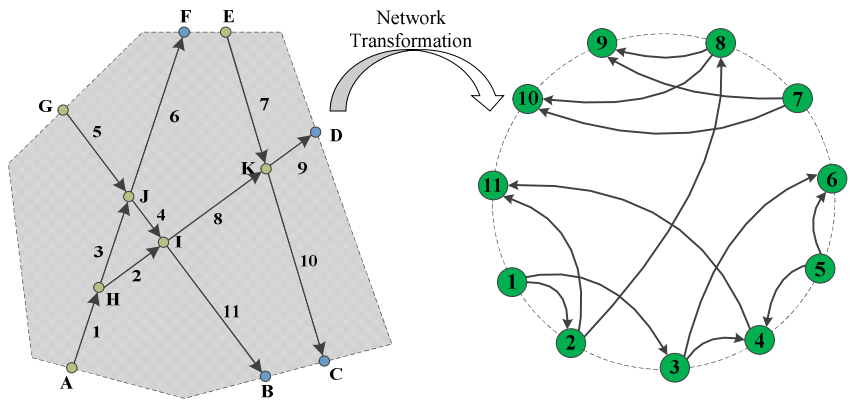

Figure 3. The schema of network dualization.

Comparing to complex networks like social network and communication network, terminal airspace networks have more regular and simple structures with relatively lower clustering coefficients[2] due to principled flight procedure design.

---

[2] Clustering coefficient is a measure of the degree to which nodes in a graph tend to cluster together. The usual definition of clustering is

The in-degrees and out-degrees of the RAN are among the key attributes of the airspace structures; others include:
- **Spatial attributes**, which includes, for each vertex in the dual network (i.e. edge in the original network), the longitudes, latitudes and altitudes of the head and tail vertices in the original network.
- **Origin & destination (OD) attributes.** A vertex is deemed an origin if its in-degree is zero, and a destination if the out-degree is 0. An origin and destination vertices are the start and end of a *standard flight path*, respectively.

## 2.2. Flight Trajectory Network

The FTN is formed by historical/empirical flight trajectories recorded by radar. It encapsulates the temporal-spatial distribution of air traffic in the terminal airspace, and provides essential information for constructing the IFDN. Each trajectory is a time series of a 5-tuple $(Call\_s, x, y, z, v)$ representing, respectively, the call sign, longitude, latitude, altitude, horizontal velocity and heading of a flight at different time stamps. The FTN is denoted $G_{FTN}(V_{FTN}, E_{FTN})$, where $V_{FTN}$ is the set of vertices, each representing one radar snapshot of a single flight; $E_{FTN}$ is the set of unweighted and directed edges. Two vertices are connected if and only if they have the same call sign, and the tail node is the next snapshot in the time series relative to the head node (see Figure 2). Corresponding to the standard flight paths in the RAN, we define the paths in the FTN *operational flight paths*. Naturally, there is a one-to-one correspondence between the operational flight paths and the flights.

The RAN and FTN combined provide much insight into the flow dynamics in the terminal airspace. However, no information regarding the air traffic controllers' behavior and communication patterns is accounted for. In order to analyze the "human-flow" dynamics from a network perspective, a flow-driven dynamic network is modelled by integrating RAN and FTN as detailed in Section 2.3.

## 2.3. Integrated Flow-driven Dynamic Network (IFDN)

IFDN is a time-varying network, denoted by $G_{IFDN} = (V_{IFDN}, E_{IFDN}; \tau)$, where each vertex corresponds to an edge in the conventional airspace network (or a vertex in the RAN). In particular, each vertex is a 4-tuple $(\overline{N}(\tau), \bar{\rho}(\tau), \bar{v}(\tau), \bar{q}(\tau))$ representing, respectively, average traffic volume, average equivalent density, average equivalent velocity, and average equivalent flow during time window $\tau$. These quantities will be detailed in Section 3.1. $E_{IFDN}$ is a time-dependent set of directed and weighted edges; the weight is the flow $Q(\tau)$ transmitted between two vertices (links in the physical network) during time period $\tau$.

In order to construct the IFDN, one needs to map the flight trajectories (FTN) onto the airspace network (RAN). From a physical point of view, the relationship between the vertex $\aleph_i \in V_{RAN}$ and $u_j \in V_{FTN}$ implies that the aircraft with the same call sign as $u_j$ was flying along the route segment $\aleph_i$ at the time corresponding to $u_j$. However, due to air traffic control strategies and system uncertainties, the actual trajectories often deviate from the standard flight paths as shown in Figure 4. While projecting an en-route point along the trajectory to the nearest route segment has been widely adopted in practice, this method could lead to erroneous mappings such as the one illustrated in Figure 4. In this figure, $u_{j+2}$ would be mapped to $\aleph_{i+1}$ following the minimum-distance rule as $d_1 > d_2$, while in fact the continuity of the trajectory dictates that $u_{j+2}$ should be mapped to $\aleph_i$. In order to remedy such a deficiency, this paper proposes a more robust algorithm that maps the flight trajectories to the network edges, thereby establishing connections between the FTN and the RAN, while maintaining the consistency of the flight routes. This algorithm is detailed in the Appendix.

---

related to the number of triangles in the network (Cohen et al., 2010).

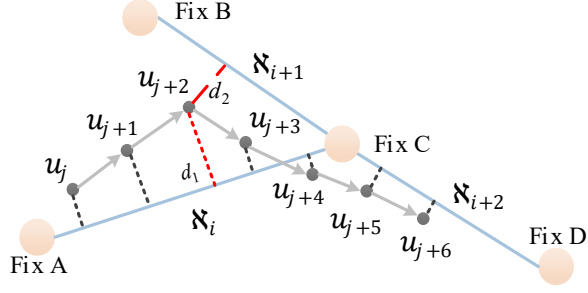

Figure 4. Schema of flight trajectory deviations. The distance between a vertex $u_i \in V_{FTN}$ and a vertex $\aleph_j \in V_{RAN}$ is indicated as the lengths of the dashed line segment, i.e. the Euclidean distance from a point to a line segment.

In the following, we illustrate the procedure for constructing the IFDN by generating cross-network edges. Links between the RAN and FTN, denoted as $E_{R-F} = \{(u_i, \aleph_j) | u_i \in V_{FTN}, \aleph_j \in V_{RAN}\}$, are the key to modeling flow-driven dynamic network. Such linkage is established as follows.

**Definition 1. (Distance between RAN and FTN vertices)** Consider a vertex $\aleph_j \in V_{RAN}$, which is a line segment expressed as an equation $Ax + By + C = 0$ in the 2-D plane. The distance between $u_i = (x_i, y_i) \in V_{FTN}$ and $\aleph_j$ is defined to be the minimum distance from the point $u_i$ to the line segment; more precisely, $dist(u_i, \aleph_j) \doteq |Ax_i + By_i + C|/\sqrt{A^2 + B^2}$.

**Definition 2. (Cross-layer connectivity)** Let $P \in \mathcal{P}_{FTN}$ and $R \in \mathcal{P}_{RAN}$ be paths in the FTN and RAN, respectively. A linkage between $u_i \in P$ and $\aleph_j \in R$ is established if and only if

1. The similarity between $P$ and $R$ is maximum, i.e. $R = \text{argmax}_{S \in \mathcal{P}_{RAN}}[simLCS(P, S)]$, where the similarity measure $simLCS(,)$ is defined in Appendix A.2.2.
2. $\aleph_j = \text{argmin}_k dist(u_i, \aleph_k)$, i.e. $\aleph_j$ is the closest to $u_i$.
3. The projection point of $u_i$ on to $\aleph_j$ lies on the route segment $\aleph_j$.

The linkages between RAN and FTN demonstrate the temporal aggregation of radar trajectory points on each link, and form star-like subnetworks centered around the RAN vertices, as shown in Figure 5; also see Figure 2 between the 2nd and 3rd layers. Since each vertex of the FTN is time-stamped, the IFDN can be also seen as a time-based integration of these star networks with the weights of the edges representing flow transitions $Q(\tau)$ between two RAN-nodes (or two flight route segments).

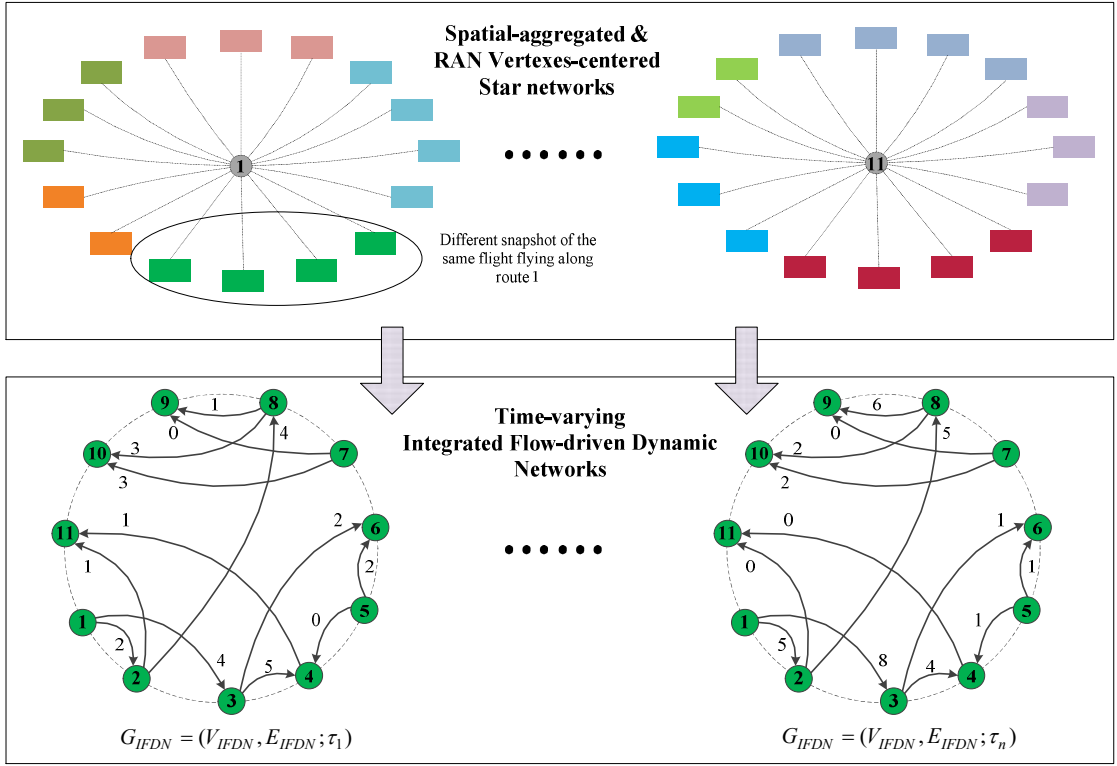

Figure 5. Integrated Flow-driven dynamic network generation based on vertex mapping between RAN and FTN

## 2.4. Interrelated "Conflict-Communication" Network (ICCN)

"Centralized Controllability" is a defining feature of air traffic flow operation, in contrast to other transportation systems. As the core of tactical management of airspace and traffic flow, air traffic control plays a vital role and represents key links in the air traffic networks studied here. In practice, the responsibility of air traffic controllers (ATCOs) is to guide aircraft through sectors in a safe, orderly and efficient way. Therefore, the control behavior, which can be regarded as a close-loop decision making consisting of the processes of monitoring, evaluation, plan formulating, and command issuing via voice and/or data link, not only plays a vital role in the evolution of traffic flows but also reveals the cognitive strategy of the ATCOs to cope with traffic complexity.

In order to study the human dynamics in air traffic control, we divide ATCOs' behaviors into two categories: internal (cognitive) complexity and external (communication) activity. ATCOs' cognitive complexity is the key attribution that reflects the difficulty in comprehending and predicting air traffic flow situation (Hilburn, 2004). Potential aircraft conflict, which marks the degree of flow disorder, is proven to be the most significant factor that influence ATCOs' cognitive complexity and workload. Regarding the external activity, air-ground communication is the integrated output of human cognition (Wang et al, 2016). Inspired by these relevant studies, we model the ICCN as a two-layer network $G_{ICCN}(V_{ICCN}, E_{CS}, E_{IC}; \tau)$ consisting of sub-networks of (1) conflict situation and (2) air-ground communication, which share the same vertices at the sector level.

### 2.4.1. First layer: conflict situation network（CSN）

Potential flight conflicts are related to the continuous approaching (i.e. the trends of flights' relative movements) that result in the loss of separation. It should be noted that flight conflicts under "ATCO-centered" terminal airspace operation have their own uniqueness. Firstly, potential conflicts are sector-based, i.e. only conflicts in the same sector are considered. Secondly, the separation between aircraft is determined by not only safety standard but also the required separation at fixes and sector boundaries. As a result, unlike en-route operation, aircraft at different flight levels are often required to maintain a certain horizontal separation in terminal airspaces, especially when they have the same destination. Therefore, we define the potential conflict in terms of the possible violation of expected horizontal separation.

Since conflict is dynamic, interpreted on a spatio-temporal domain, conflict situation is modelled as a time varying network $G_{CS}^X(V_{ICCN}(\tau), E_{CS}(\tau))$, where $V_{ICCN}(\tau) = \{f_1, f_2, \ldots, f_C\}$ is a finite set of vertices representing existing aircraft during time period $\tau$ in the sector X; $|V_{ICCN}(\tau)| = C$ is the total aircraft count; $E_{CS}(\tau)$ is a finite set of non-directional and unweighted edges. At any time instance $t$ during the window $\tau$, the sets of positions and speeds are respectively $\{U_i = (x_i, y_i): i = 1, \ldots, k\} \subset \mathbb{R}^2$ and $\{V_i = (\hat{x}_i, \hat{y}_i): i = 1, \ldots, k\} \subset \mathbb{R}^2$. Then, the relative position and speed are $U_{ij} \doteq U_j - U_i$, $V_{ij} \doteq V_j - V_i$, respectively. We say that $i$ and $j$ have potential conflict if and only if there exists $\Gamma_{ij} \in \mathbb{R}_+$ such that

$$\begin{cases} \|U_{ij} + \Gamma_{ij} \cdot V_{ij}\| \leq S \\ \Gamma_{ij} \leq \min(t_i, t_j) \end{cases} \quad (1)$$

where $S$ is the expected separation, $t_i$ and $t_j$ are the remaining flight time of aircraft $i$ and $j$ in sector X, respectively. (1) simply says that flights $i$ and $j$ are in potential conflict in sector X if, by the time that either one of them leave the sector X, their relative distance is within the expected separation $S$.

*2.4.2. Second layer: interventional communication network (ICN)*

Under current mode of air traffic control operation, air-ground communication is the primary means to issue control commands, and is regarded as the integrated output of the ATCOs' mental processes. It is important to understand its basic patterns and underlying mechanism. Motivated by the recent findings that human communication and other interactive activities exhibit characteristics of heavy-tailed, power-law distributions, a similar data-driven approach was adopted (Wang et al., 2016) to investigate intercommunication events to characterize temporal-spatial behavior of controller communications. General analyses on communication activities give rise to emerging spatial patterns of human dynamics. In order to further understand the underlying factors that influent vocal communication patterns and human cognition, here we define an Interventional Communication Network (ICN) to illustrate the temporal-spatial characters of active communication behavior.

ICN is a weighted non-directional time-varying network denoted as $G_{ICN}^X(V_{ICCN}, E_{ICN}; \tau)$. The vertices coincide with those of the conflict situation sub-networks, and represent all the flights in the sector X during time window $\tau$. The edges record the frequency of highly related interventional communications. The interventional communication is defined as the Controller-to-Pilot (C-P) commands that aim at adjusting flight status (e.g. speed, altitude and heading) in order to improve flow efficiency or to avoid potential conflicts. The following definition is adopted from Wang et al. (2016).

**Definition 3. (Temporal Distance)** Given flights $i$ and $j$, their temporal distance at time $t$ is defined as $D(i, j; t) = s_j(t) - s_i(t) - l_i(t)$, where $s_i(t)$ denotes the start time of the latest call event for flight $i$, and $l_i(t)$ is the duration of this call. This is illustrated in Figure 6.

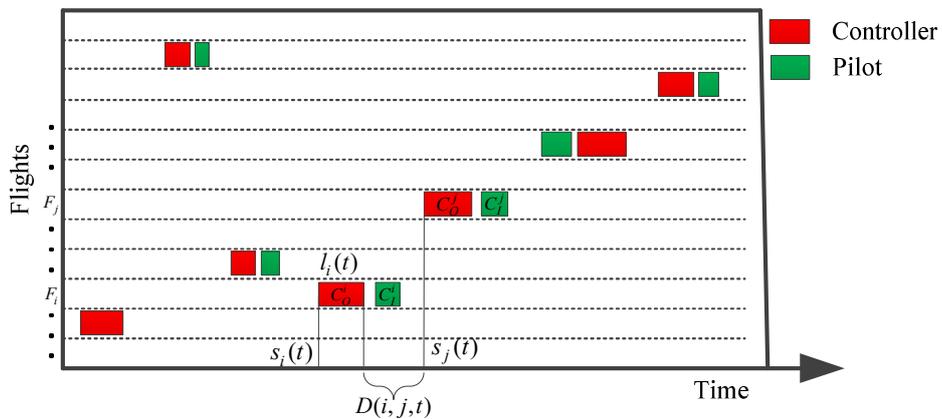

Figure 6. Temporal communication activity

**Definition 4. (Connectivity in the Temporal Network)** Flights (vertices) $i$ and $j$ are connected at time $t$ if $D(i, j, t) \leq t_{min}$, and the type of command is "Interventional Communication", where $t_{min}$ is the threshold of Temporal Distance.

During time window $\tau$, the weight of an edge in the ICN is the number of times of observed connectivity; see Wang et al.

(2016). The ICN changes from one time windows $\tau$ to another. It captures the relationships among flights that are determined by the communication initiated by controllers and pilots.

## 2.5. Summary

The primary aim of this paper is to explore the complex dynamics in air traffic and activities of air traffic controllers (ATCOs) in an empirical and comprehensive way. To achieve this, air traffic flow dynamics, "ATCO-flow" interactions, and system-level non-linear chaotic dynamics are investigated in detail using different network representations of such complex dynamics. The multi-layer network is proposed as an effective tool to integrate source data and analytical metrics, and consolidate interpretation of traffic and ATCO dynamics. It is not the primary focus of this paper to establish a multi-layer network with coherent relationships among all its components, without empirical exploration of the complex dynamics.

A few structural relationships exist among the physical and functional networks within the multi-layer network. First of all, IFDN is a dynamic network that presents the flow transmission in the airspace network. As described in Section 2.3, to model IFDN, generating the cross-layer linkage between RAN and FTN is crucial to identify the aircraft flying along each route segment (see Figure 5), and to derive network flow metrics stated in Section 3.1. Secondly, we develop the ICCN based on the integrated RAN-FTN network and controllers' communication data. For the ICCN of each sector, the vertices, which denote the aircraft in the sector during time period $\tau$, is extracted from FTN with the help of cross-layer linkage between RAN and FTN. The edges of CSN are modeled based on inheriting attributes (position, speed) of the FTN vertices to calculate the potential conflicts.

# 3. Analytical Metrics

## 3.1. Network flow metrics

Flow rate, density and velocity are fundamental dynamic parameters. Unlike road vehicle operations, air traffic in controlled airspace presents unique characteristics influenced by ATCOs. When congestion occurs in the terminal airspace, maneuvering commands, including speed reduction, detour, shortcut, and hold are issued to change pre-planned dynamic trajectories to avoid potential conflicts. In order to accommodate the unique feature of air traffic flow in terminal areas, which often deviate from standard flight routes, we propose a 4-tuple $(\overline{\mathbb{N}}(\tau), \bar{\rho}(\tau), \bar{v}(\tau), \bar{q}(\tau))$ for the vertices of the IFDN. For one time window $\tau$, we are given the time stamps $t_i$ used in the OFP, $i = 1,2,\ldots,N$. The cross-network degree (i.e. between the RAN and FTN) of $\aleph \in V_{RAN}$ at time $t_i$ is $M(t_i)$; that is, $M(t_i)$ represents the number of aircraft traveling along route segment $\aleph$ at time $t_i$.

(1) **Average Traffic Volume** (ATV) $\overline{\mathbb{N}}(\tau)$. The ATV of a route segment (expressed as a vertex in the RAN) $\aleph$ is formulated as:

$$\overline{\mathbb{N}}(\tau) \doteq \frac{1}{N}\sum_{i=1}^{N} M(t_i) \qquad (2)$$

(2) **Average Flow Rate** (AFR) $\bar{q}(\tau)$. The AFR is the average outflow from $\aleph \in V_{RAN}$ during time period $\tau$ with length $|\tau|$. Let the sum of weights of outgoing edges of vertex $\aleph$ in the IFDN be $Q(\tau)$, then $\bar{q}(\tau) = Q(\tau)/|\tau|$.

(3) **Equivalent Average Density** (EAD) $\bar{\rho}(\tau)$. For any route segment $\aleph \in V_{RAN}$, its start and end points are denoted $\aleph^s$ and $\aleph^e$, respectively. For each $t_i$, the locations of the flights connected to $\aleph$ are $u_m(t_i)$, $m = 1,\ldots,M(t_i)$. Then the average density at time $t_i$ is defined as (here, all nodes are interpreted as points on a two-dimensional plane):

$$\bar{\rho}(t_i) \doteq \begin{cases} \dfrac{M(t_i) - 1}{\max_m \|u_m(t_i) - \aleph^e\| - \min_m \|u_m(t_i) - \aleph^e\|}, & \text{if } M(t_i) > 1 \\ \dfrac{M(t_i)}{\|\aleph^s - \aleph^e\|}, & \text{if } M(t_i) \leq 1 \end{cases} \qquad (3)$$

The EAD is therefore defined as the time average:

$$\bar{\rho}(\tau) \doteq \frac{1}{N} \sum_{i=1}^{N} \bar{\rho}(t_i) \tag{4}$$

Compared to the conventional definition of density, which is the ratio of vehicle count to segment length, EAD highlights the operational characteristics of route-based maneuvering in terminal airspace.

(4) **Equivalent Average Speed** (EAS) $\bar{v}(\tau)$. For each $t_i$, the speeds of the flights connected to $\aleph$ are denoted $v_m(t_i)$, $m = 1, \dots, M(t_i)$. EAS is the average effective speed of aircraft travelling along the route segment; behaviors like detour, shortcut and holding in congestion situations can be characterized by the fluctuation of EAS.

$$\bar{v}(\tau) \doteq \frac{1}{N} \sum_{i=1}^{N} \frac{1}{M(t_i)} \sum_{m=1}^{M(t_i)} v_m(t_i) \cdot \beta^m \tag{5}$$

where $\beta^m$ is the Velocity Gain Coefficient corresponding to $\aleph$ and flight $m$, which is defined to be the ratio of the standard route length to the actual flying distance.

## 3.2. Complexity metrics of ATCOs

### 3.2.1. Solution space-based perceived complexity (SSPC)

Certain horizontal separation should be maintained for aircraft in the terminal airspace. Speed and heading vectoring is the primary means to avoid conflict and keep required headway in controlled flows. The *solution space* of potential conflict is defined as the 2-D area of continuous heading-and-speed space (d'Engelbronner et al., 2015). In practice, however, the solution space changes with solution strategy (e.g. sequencing) and restricted by airspace structure and flight procedure (e.g. heading change limitations). In order to reflect such a reality, we propose a complexity metric for conflict situation networks based on modified shadow-based solution space to reflect the urgency and difficulty of potential conflicts.

We assume that aircraft $A$ and $B$ are in potential conflict as shown in Figure 7. Limited by air traffic control regulations and aircraft performance, the heading and speed maneuvering spaces of $A$ at time $t_i$ are $[\theta_{A1}, \theta_{A2}]$ and $[v_A^{\min}, v_A^{\max}]$ respectively. We define the area of the Total Maneuvering Space (colored by gray in Figure 7 (a)) of aircraft A to be

$$TMS_A(t_i) = 0.5(\theta_{A2} - \theta_{A1})((v_A^{\max})^2 - (v_A^{\min})^2).$$

Then, the combined solution spaces of aircraft A under different sequencing strategies are colored by brown as shown in Figure 7.

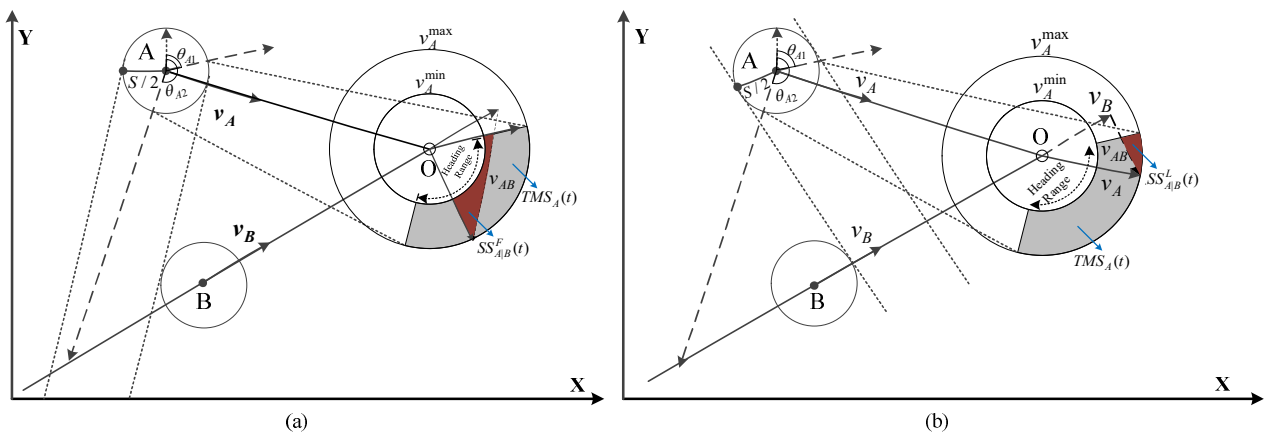

Figure 7. Scheme of solution space of aircraft A in potential conflict. (a) Solution space in the condition of aircraft A follows B; (b) Solution space in the condition of aircraft B follows A.

It is easy to prove that there is no intersection of solution space under these two ordering (sequencing) strategies. So, the total solution space of aircraft $A$ is the sum of solution space in leading and following conditions, denoted as $SS_{A|B}(t_i) = SS^L_{A|B}(t_i) + SS^F_{A|B}(t_i)$. Likewise, the total solution space of aircraft $B$ can be derived similarly. The average solution space area of aircraft $A$ during time period $\tau$ is denoted as $SSA_A(\tau) = \sum_{i=1}^{N_A} \Lambda\left(SS_{A|B}(t_i)\right)/N_A$, where $N_A$ is the number of snapshots of aircraft $A$ during time period $\tau$, and function $\Lambda(\cdot)$ is the area operator.

In high-density air traffic operation, multi-aircraft conflicts emerge commonly. Assuming aircraft $A$ is in conflict with more than one aircraft, the urgency index is formulated as $SSA_A(\tau) = \sum_{i=1}^{N_A} \Lambda\left(\cap SS_{A|X}(t_i)\right)/N_A$, where X is the aircraft that have conflict with A; $\cap SS_{A|X}$ is the intersection of solution spaces that can simply be calculated using the method in Figure 7.

The solution space approach not only provides dynamic pictures of microscopic structure of air traffic flow, but also promises novel ATCO's cognition complexity measurements. More intuitively, Solution Space-based Perceived Complexity $SSPC(\tau)$ is modelled as the average difficulty of finding solution space for aircraft in the same sector during time period $\tau$ based on the relative area of solution space.

$$SSPC(\tau) = \frac{1}{\mathbb{Q}(\tau)} \sum_{j=1}^{\mathbb{Q}(\tau)} \left(\left(SSA_j(\tau)/TMS_j(\tau)\right)^{-\sigma} - 1\right) \quad (6)$$

where $TMS_j(\tau) = \sum_{i=1}^{N_A} TMS_j(t_i)/N_A$ and $\mathbb{Q}(\tau)$ are the average Conflict-free Solution Space of aircraft $j$ and the number of aircraft emerged in the sector during time period $\tau$ respectively; $\sigma$ is a co-efficient, which is simply set as -0.5 here, describing the non-linear impact of solution space on cognition complexity. The value range of $SSPC$ value is $[0, +\infty]$. If and only if there is no conflict during some time period, $SSPC$ equal to 0.

*3.2.2. Communication load (CL)*

Communication load is the primary composition of ATCOs' workload (Porterfield et al., 1997). Study on the adjustment of communication behavior may reveal controllers' metacognition dynamics to cope with complexity. Communication load is defined as the percentage of air-ground communication channel occupancy in certain time period as shown in formula (7).

$$CL_s(\tau) = \left(\sum_{i=1}^{I} \beth_i^{c \to p} + \sum_{j=1}^{J} \beth_j^{p \to c}\right)/\tau \quad (7)$$

where $\beth_i^{c \to p}$ and $\beth_j^{p \to c}$ are the duration of the $i^{th}$ controller-pilot and $j^{th}$ pilot-controller communication respectively.

### 3.3. Non-linear analysis metrics

Chaotic analysis is a modern tool for identifying the high-level characteristics of non-linear dynamic systems (Abarbanel et al., 1993). The dynamics of "human-flow" interacted air traffic system is always regarded as complex and nonlinear, and cannot be described using a group of functions. Shi pointed out that the first-line task of fully achieving automated air traffic management is to figure out the complex chaotic problems fall in between randomness and certainty (Shi, 2001) of air traffic system. Given continuous data series of system variable $\mathcal{H}_s^\tau = [h(t_1), h(t_2), \ldots\ldots, h(t_n)]$, the identification of chaos mainly includes two part: reconstruction of system phase space and chaos proof.

*(1) Reconstruction of phase space*

For complex system, it would be very difficult to develop a perfect model to describe the system behavior. In most cases, only time series of one or more variables of the system can be measured. Conventionally, time series is analyzed in time domain or transform domain. However, to deal with chaotic time series, all calculations are implemented in the phase space which is an approximation of real-world system. The theory of reconstruction of phase space points out the evolution of any component is determined by the interactions with other components in the system and can be reproduced by optimal time delay and embedded dimension (Parked et al., 1980; Takens et al., 1981) as shown is formula (8).

$$H = \begin{bmatrix} h(t_1) & \cdots & h(t_M) \\ \vdots & \ddots & \vdots \\ h(t_1 + (m-1)\tau) & \cdots & h(t_M + (m-1)\tau) \end{bmatrix} \quad (8)$$

where $M$ is the number of phase point in $m$-dimensional phase space, $M = n - (m-1)\tau$, $\tau$ is the delay time.

*(2) Chaos identification algorithm*

Lyapunov exponent, as a key metric to measure system dynamics, characterizes the rate of divergence of nearby trajectories in the phase space. The Largest Lyapunov Exponent (LLE) is a notion of predictability for a dynamical system. A positive LLE is usually taken as an indication that the system is chaotic (Gaspard, 1998).

These methods of phase space reconstruction and chaos identification have been widely used in chaos analysis of natural, social, and sociotechnical systems. Here, only brief descriptions of the process are provided as shown in Figure 11. Detailed methods adopted refer to chaotic analysis in Cong et al. (2014).

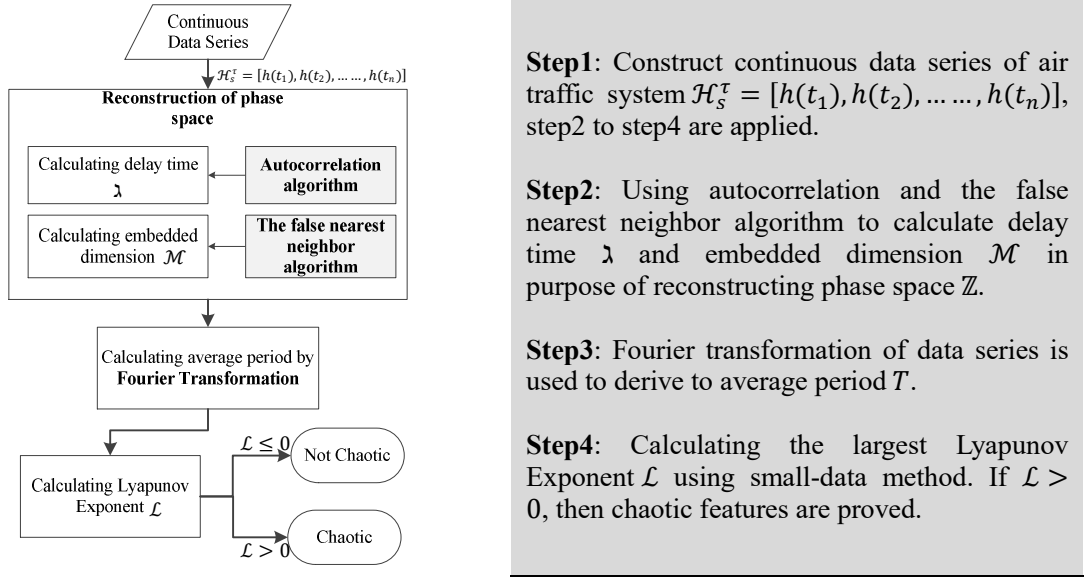

**Step1**: Construct continuous data series of air traffic system $\mathcal{H}_s^\tau = [h(t_1), h(t_2), \ldots \ldots, h(t_n)]$, step2 to step4 are applied.

**Step2**: Using autocorrelation and the false nearest neighbor algorithm to calculate delay time $\lambda$ and embedded dimension $\mathcal{M}$ in purpose of reconstructing phase space $\mathbb{Z}$.

**Step3**: Fourier transformation of data series is used to derive to average period $T$.

**Step4**: Calculating the largest Lyapunov Exponent $\mathcal{L}$ using small-data method. If $\mathcal{L} > 0$, then chaotic features are proved.

Figure 8. Process of calculating the largest Lyapunov Exponent

# 4. Empirical results: Case study of Guangzhou terminal airspace

## 4.1. Data description

The Guangzhou terminal airspace is mainly responsible for the inbound and outbound traffic of Baiyun International Airport, one of the top three busiest airports in China. In order to explore the dynamic features of the air traffic in the terminal airspace based on the multi-layer network framework and analytical metrics, synchronized trajectory data and air-ground communication data on three typical days 15/05/2012, 11/09/2012 and 18/12/2012 are collected. The trajectory data have a time resolution of 15 seconds. The airspace configuration and sample trajectory data are shown in Figure 9. The communication data contain records in the radio channel for each sector, including the start and end times of each voice communication. The sample time series of air traffic volume and communication activities are shown in Figure 10.

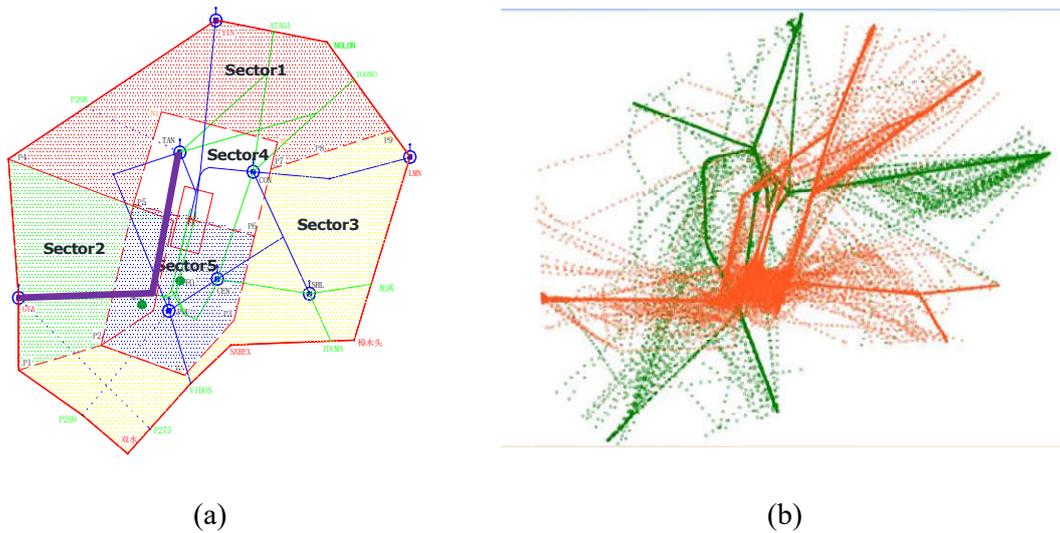

Figure 9. Trajectory data in Guangzhou terminal airspace. (a) Airspace configuration; the segments highlighted with purple are the merge routes GYA-AGVOS and TAN-AGVOS. (b) Trajectory data of arrival (Orange) and departure (Green).

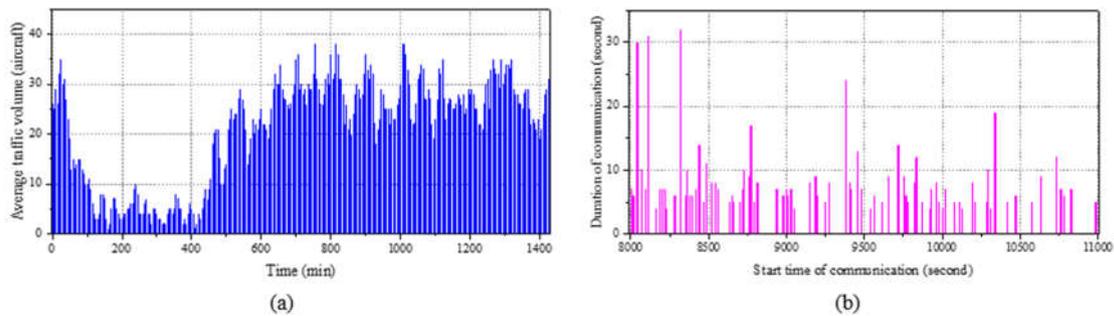

Figure 10. Sample data of traffic volume and communication activities. (a) Average Traffic volume per 5 min in terminal airspace on 11/09/2012; (b) Duration of each communication initiated by air traffic controller in Sector 2.

## 4.2. The fundamental diagrams of flow dynamics

### 4.2.1. Route-based Fundamental Diagram

The Fundamental Diagram (Daganzo, 1994) presents a one-to-one relationship between aggregated road traffic flow variables (e.g. flow and density), and exhibits phase transitions on a road segment. Due to the uneven and sparse spatial distribution of air traffic flow, it is difficult to observe the complete phases on a single air route. In order to remedy such a situation, we choose the busiest merge routes GYA-AGVOS and TAN-AGVOS highlighted in Figure 9(a). The traffic on the merge routes accounts for nearly 50% of the total arrivals. The IFDN provides a dynamic picture of flow transmission among route segments by consolidating RAN and FTN of Guangzhou Terminal Airspace.

We set the length of the time window $\tau$ in Section 3.1 to be 5 min. Figure 11(a) shows the non-linear relationships among Average Flow Rate (AFR), Equivalent Average Density (EAD) and Equivalent Average Speed (EAS). Generally, as EAD increases, more intense constraints among aircraft lead to decreased EAS to varying degrees, while the EAD-EAF relationship exhibits a concave shape with close-to-linear increase of the flow for relatively low density. These observations reveal similar patterns in road traffic. For detailed analysis of route-based flow dynamics, four phases are divided by dissecting both fundamental diagram (Figure 11(c)-(d)) and temporal-spatial diagram (Figure 11(b)) with the additional help of flight data replay using SIMMOD simulator.

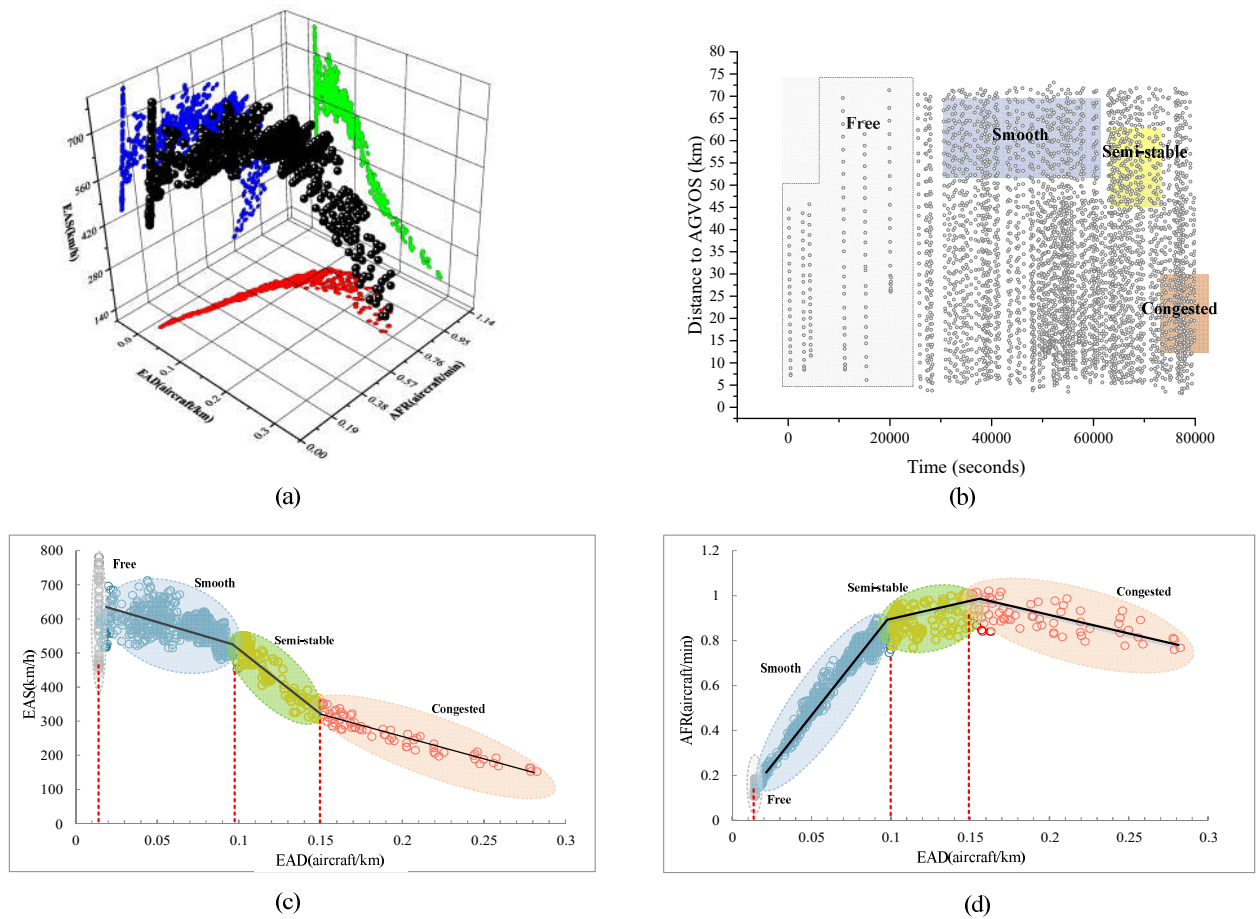

Figure 11. Typical fundamental diagram configuration of route-based air traffic flow. (a) The EAD-EAF-EAS relationship; (b) Temproal-spatial diagram of air traffic flow along merge routes GYA-AGVOS and TAN-AGVOS; (c) Basic phase state implied in flow evolution presented by EAD-EAS relationship; (d) Basic phase state implied in flow evolution presented by EAD-EAF.

(1) **Free Phase** corresponds to extremely low flow density, large spatial headways and little interactions among aircraft. For the different speed profiles assigned to flights based on individual performance, the EAS presents significant fluctuation. As shown in Figure 12(a), shortcut strategies observed by flight distance reductions and high velocities indicated by the steep trajectories increase the flow efficiency.

(2) **Smooth Phase** corresponds to relatively high EAS with a slight reduction compared to the Free Phase due to occasional conflicts. EAF increases linearly as EAD grows. In this phase, aircraft are lined up in standard flight route with approximately equal flight distances as observed in Figure 12(b), although the spatial headways are not evenly distributed.

(3) **Semi-stable Phase** where aircraft are still flying along standard route with smaller and more uniform spatial headways as shown in Figure 12(c). A discernible decrease in the EAS is observed due to rising conflicts resolved mainly with speed reduction and occasional heading change. The EAF achieves maximum range with a slower increment when EAD increases. The utilization of airspace resource has peaked, and the traffic is not stable, on the edge of phase transition when disturbed.

(4) **Congested Phase** where both EAS and EAF drop as EAD continues to increase. From an operational point of view, the resolution of conflict switches from speed strategy to radar vectoring and holding as shown in Figure 12(d). From a macroscopic perspective, the streamline of traffic changes from structured linear state into disseminative planar state. The situation of air traffic falls into chaos and out-of-order. The limitation of the ATCOs in calculating parameters of radar vectoring and holding (e.g. outbound angle, holding time) is also a primary reason leading to inefficient usage of capacity.

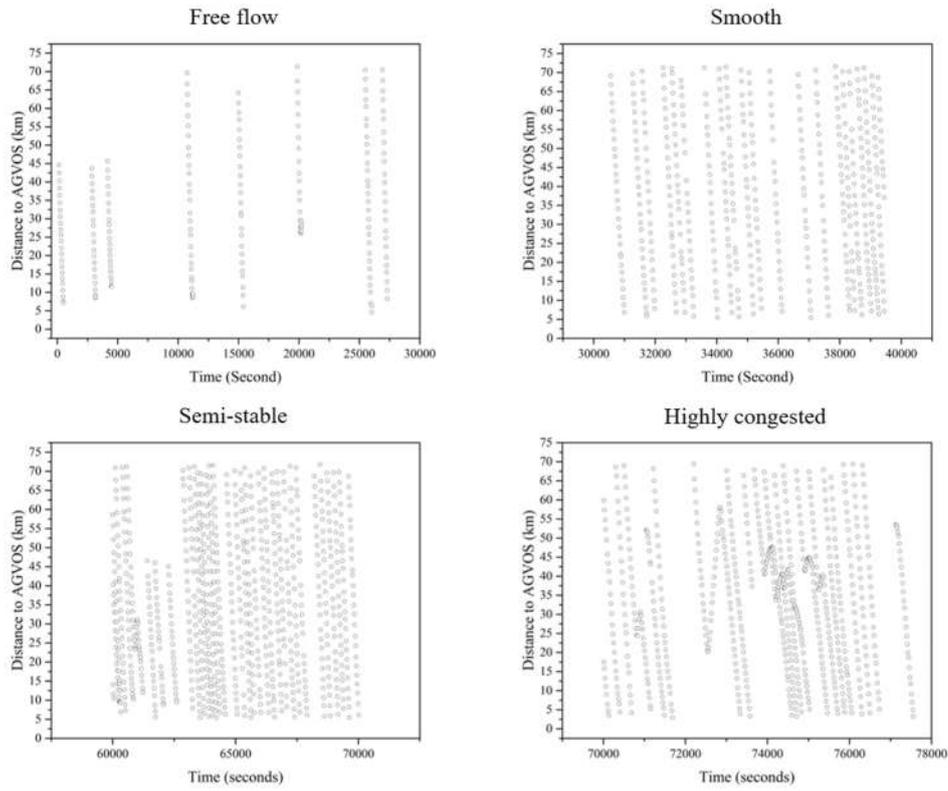

Figure 12. Temporal-spatial diagram of traffic flow in merging routes

### 4.2.2. Network-based Macroscopic Fundamental Diagram

In most terminal airspaces, arrival and departure air routes are physically or operationally isolated by lateral or vertical separations. In the case of Guangzhou, only one route segment (SHL-IDUMA) is laterally shared by arrival and departure flows but vertically separated. Therefore, the runway is the primary point of interaction and conflict between arrival and departure flows. Given the little constraint that arrival and departure traffic impose on each other in the airspace, we divide the terminal airspace network into arrival network and departure network, which are linked by the runway; see Figure 13(a). The star networks centered at RAN vertices (recall Figure 5) are shown in Figure 13(b).

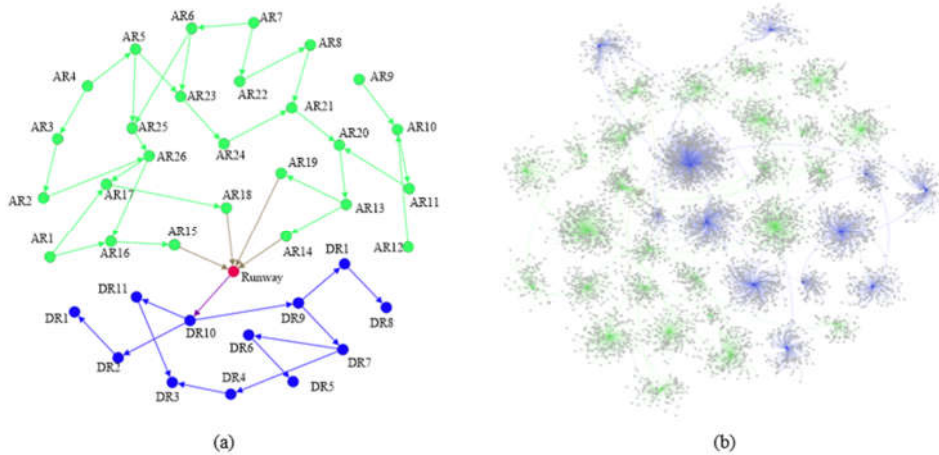

Figure 13. (a) Route-based Airspace Network; green: arrival network, blue: departure network. (b) Spatial-aggregated &RAN Vertices-centered Star Networks.

Unlike the fundamental diagram, the macroscopic fundamental diagram (MFD) (Geroliminis et al., 2008) characterizes the aggregate behavior of networked traffic in term of occupancy and throughput, in a parsimonious way capable of capturing the key demand-supply relationship. Here, in deriving the MDF, we follow Section 3.1 and construct the various network flow variables. Given the average flight time in the terminal airspace and conventional time interval for statistical analyses in air traffic research, we set the length of the time window $\tau$ to be 15 min. For the reason stated above, we distinguish between MFD-Arrival and MFD-Departure, defined based on the arrival RAN and departure RAN, respectively.

Figure 14(a) shows the MFD-Arrival, representing the relationship between average arrival demand and average runway arrival rate. The bars show the average while the shared area indicate standard deviation. The Pearson correlation test indicates a high positive correlation of 0.827 between arrival demand and arrival rate; the two-tailed probability passes the significance level test with a p-value of 0<0.01. It is widely accepted in air traffic operation that the system throughput continuously rises with increased demand. However, our result shows a slight drop of arrival rate (throughput) past the critical arrival demand. The main reason is the reduced speed caused by airspace network congestion and expanded time headway between arrivals especially at the final approach stage. Moreover, this is also partially attributed to the limitations of ATCOs' capabilities to process complex situations, and high safety buffer in congested cases leading to low utilization of network capacity.

Figure 14(b) shows the impact of runway departure rate on MFD-A. The MFD-A is not sensitive to the increased departure rate at lower levels, which may be explained by the runway capacity envelop configuration (Gilbo, 1993). When the runway departure rate continues to increase, the average arrival rate decreases noticeably due to the competition for runway resource. Finally, the critical arrival demand that maximizes the arrival rate for a given departure rate decreases as the runway departure rate increases, as shown by the dots in Figure 14(b).

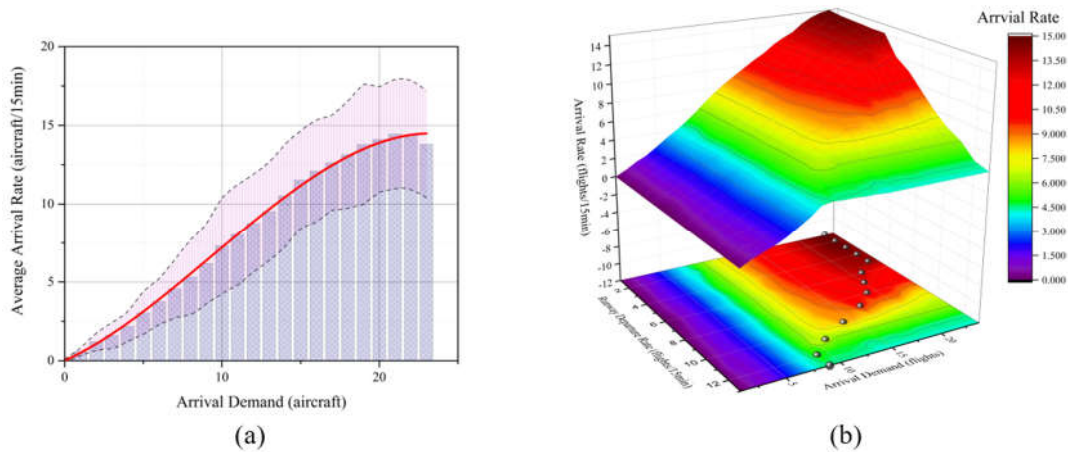

Figure 14. Macroscopic fundamental diagram of arrival network flow. (a) Average arrival rate under different demand. (b)Impact of departure rate on arrival MFD

In the departure network, the MFD-D is defined in terms of departure demand $\sum_{\aleph_n \in V_{RAN-D}} \overline{\mathbb{N}}_n(\tau)$ and departure rate. Here, departure rate is defined as the sum of outflow rate at departure fixes. Figure 15 shows little congestion effect in the MFD-D, due to controlled runway separation and diverging configuration of streamlines in general as shown in Figure 13 (a). A higher positive correlation (0.911) and significance level (p-value<0.01) between departure demand and airspace departure rate is confirmed by the Pearson correlation test. As shown in Figure 15(b), unlike the MFD-A, the impact of runway arrival rate has little effect on the shape of the MFD-D.

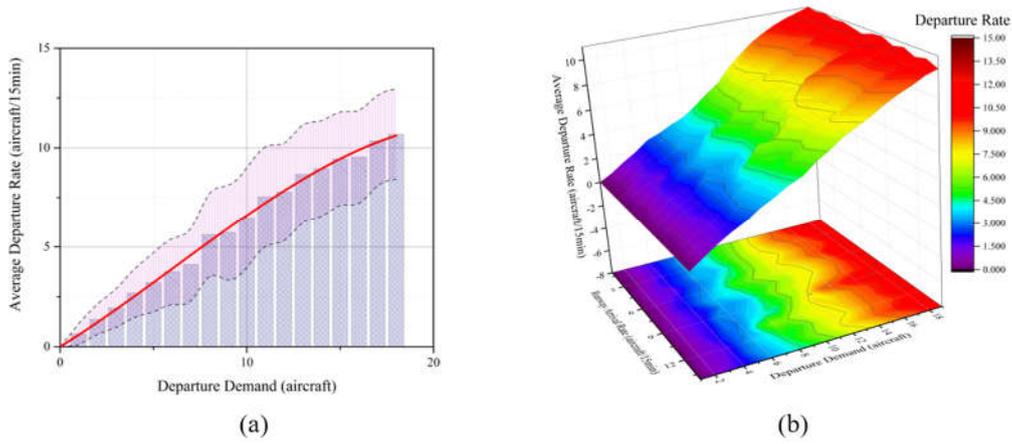

Figure 15. Macroscopic fundamental diagram of departure network flow. (a) Average departure rate under different demand. (b)Impact of arrival rate on departure MFD

FD and MFD provide aggregated view and intuitive understanding of the flow dynamics at the route and network level. They also underpin air traffic flow modeling and optimization at mesoscopic and macroscopic levels, as suggested by the plenty of studies in road traffic (Geroliminis et al., 2012, Hajiahmadi et al., 2013). However, a defining distinction between road and air traffic is the direct involvement and influence of ATCOs who are held responsible for the safety and efficiency in their own sectors. Therefore, a comprehensive analysis of air traffic dynamics in terminal areas cannot be done without incorporating the human-flow interactions, as we shall explore in the next section.

## 4.3. Underlying mechanism of "human-flow" interaction at sector level

A terminal airspace consists of sectors that are collaboratively controlled by ATCs. The ATCO's behavior can be regarded as close-loop control processes with planning, implementation, monitoring and evaluation (Hilburn, 2004). In the "ATCO-centered" air traffic tactical operation, the initial inflow pattern, output flow expectation and control behavior are the key factors that determine the evolution of air traffic flow inside the sector. In this section, we interpret the underlying mechanism of "human-flow" dynamics in airspace sectors. Here, we set time period $\tau = 15$ min.

Figure 16(a) shows sector-level MFDs obtained with 1-min time resolution. Unlike the MFD in terminal airspaces (see Section 4.2.2), sector-level MFDs tend to exhibit significant over-critical traffic, especially for high-demand sectors like 1, 2 and 5, where the average outflow rate experiences a discernible drop. In order to investigate the detailed dynamics of human-flow evolution, we additionally plot average EAS, outflow rate, CL and SSPC for all the 5 sectors, as shown in Figure 16(b)-(f), respectively.

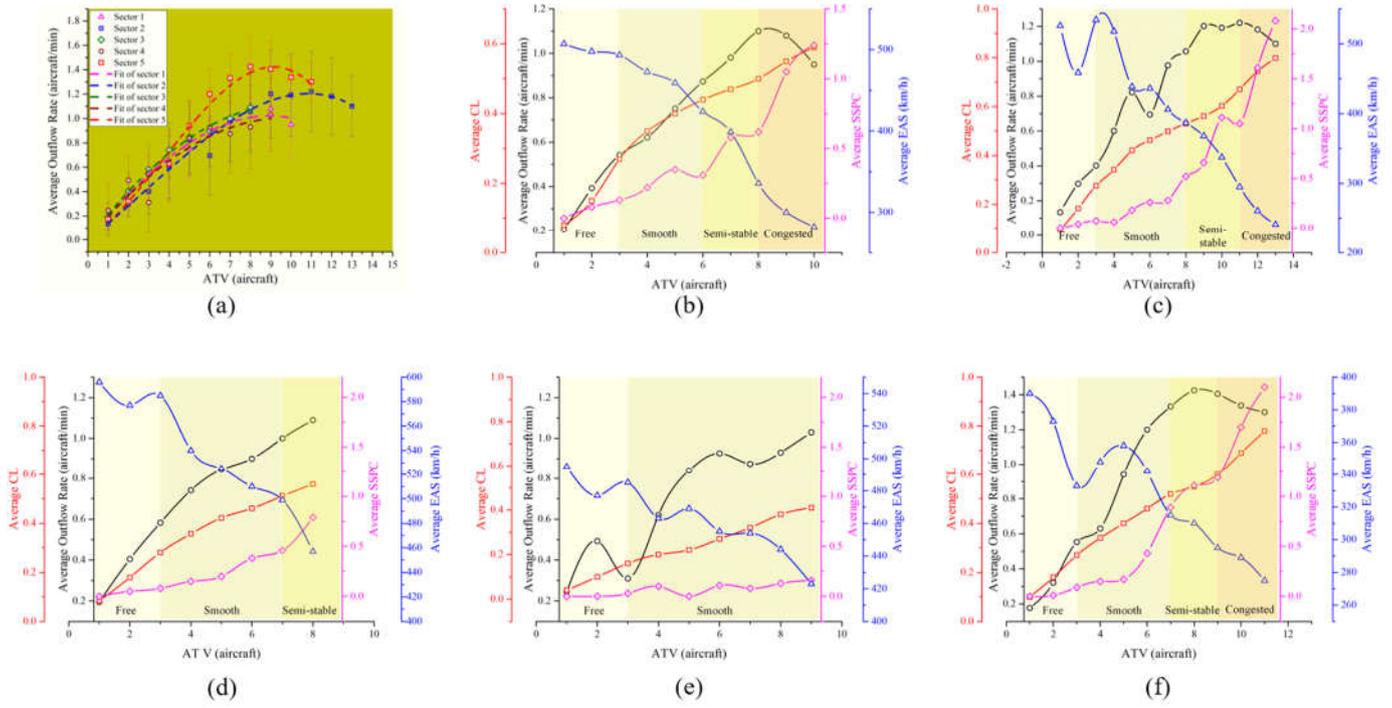

Figure 16. "ATCOs-flow" performance in Guangzhou terminal sectors.

Intuitively, increased traffic leads to more conflict, lower flow efficiency and higher workload. Correlation analysis shows that EAS has a negative correlation with ATV: average Pearson Correlation Coefficient -0.811 and p-value≈0<0.01. Communication load and SSPC show positive correlations with traffic: average Pearson Correlation Coefficient 0.851 and 0.798 respectively, both p-values≈0<0.01. Based on the discussion of phases in Section 4.2.1, we will analyze the curves in Figure 16 in conjunction with radar data, and show that the four traffic phases are also suitable to describe the sector traffic. In the following discussion we present some novel findings on ATCOs' performance, which reflect their metacognition (Flavell, 1979) dynamics. This highlights the "human-flow" interactions through both qualitative and quantitative analyses based on Figure 16 and ICCN. Specifically, we focus on the degree distribution and motifs of CSN (Section 2.4.1) and ICN (Section 2.4.2) – the two layers of ICCN – in order to uncover patterns of interactions between ATCOs and traffic situation. The motifs are closely related to the conflict structure and control decisions made by ATCOs. We define the following three basic motifs used in the study of social communication (Zhao et al., 2010): chain, loop and star (see Figure 17).

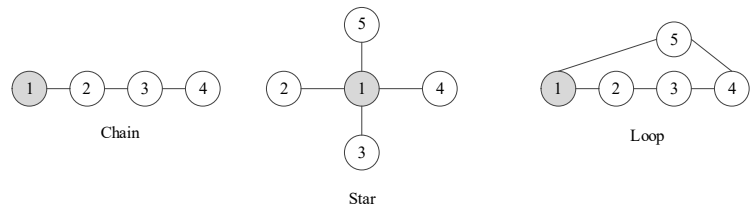

Figure 17. Basic motifs in network structure

We begin with some similarities between "ATCOs-flow" performance and ICCN attributes in each sector as the traffic phases evolves in time.

- With the increase of ATV, CL and SSPC rise simultaneously and non-linearly as shown by the fluctuant average increase rate in Table 1. Generally, CL increased much more slowly than SSPC except in the free phase due to meta-cognition dynamics explained later. Particularly, extreme low SSPCs in sector 4 are due to its simple diverging route structure with few conflicts.
- The empirical degree distributions of ICN (denoted $p_I$) and CSN (denoted $p_C$) vary by the traffic phases. As $t_{min}$ increases $p_I$ becomes less left-skewed, and its expectation increases as well. This is evident across all four phases

- and coincides with intuition.
- We observe positive correlations with moderate significance between the degrees of vertices shared by CSN and ICN except in the free phase (Figure 18(b)). The fact that degrees of vertices of two layers in ICCN present higher correlation in busier traffic situation implies that as the increase of congestion, interventional communications are mainly triggered passively by potential conflicts rather than pro-actively (e.g. trajectory adjustment) in light traffic situations. This is partially supported by the discussion of phases in Section 4.2.1.
- Chains and stars are the most common motifs in CSN, while chains and loops appear mostly in ICN. Noted that more motifs of loops in ICN emerge as the increase of traffic load and star-like motifs in CSN. It implies that controllers issue circulatory commands to aircraft one by one to ensure the safety separations among aircraft in complex conflict situation.

Table 1 Average increase rate of CL and SSPC during each phase in sectors.

| Sector | Performance Index | Free | Smooth | Semi-stable | Congested |
| --- | --- | --- | --- | --- | --- |
| Sector1 | CL | 0.085 | 0.040 | 0.047 | 0.057 |
|  | SSPC | 0.065 | 0.107 | 0.135 | 0.26 |
| Sector2 | CL | 0.091 | 0.051 | 0.059 | 0.064 |
|  | SSPC | 0.037 | 0.089 | 0.095 | 0.515 |
| Sector3 | CL | 0.098 | 0.051 | 0.068 | × |
|  | SSPC | 0.040 | 0.103 | 0.330 | × |
| Sector4 | CL | 0.060 | 0.042 | × | × |
|  | SSPC | 0.021 | 0.025 | × | × |
| Sector5 | CL | 0.085 | 0.042 | 0.049 | 0.090 |
|  | SSPC | 0.045 | 0.160 | 0.171 | 0.456 |

In the following, the distinct features are discussed to demonstrate and prove the underlying dynamics of "ATCOs-flow" interaction.

In the free phase, there is a discernible increase in the ATCOs' communication loads (see Figure 16) with little to no conflict, as is evident from the heavily left-skewed distribution $p_C$ and the motifs in Table 2. Besides, extremely low correlations between the degrees of vertices shared by CSN and ICN is observed for different $t_{min}$ as shown in Figure 18(b). Moreover, motifs analysis shows that as $t_{min}$ increases, the cumulative proportion of three motifs is getting close to 100% even though the conflict situation is quite simple. Based on these results, we define one of the metacognition dynamics as "pre-activation" of cognition complexity in low workload for the preparation for unexpected traffic situations, by increasing the traffic complexity intentionally. Driven by "pre-activation", the communication load is not generated by conflict but by issuing interventional commands like shortcut, approach with high speed, and descent delay, etc.

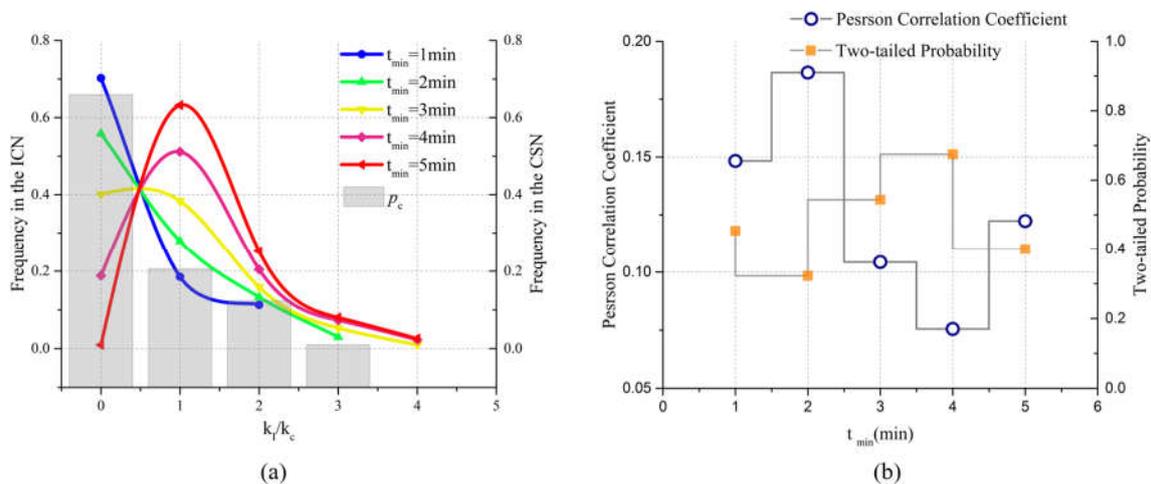

Figure 18. Free phase: (a) Degree distributions of CSN and ICN; (b) Degree Correlation for vetices shared by CSN and ICN.

Table 2 Motifs within ICCN in the free phase.

| Conflict Situation Network (CSN) | Interventional Communication Network (ICN) | | | | |
|---|---|---|---|---|---|
| | $t_{min} = 1$ | $t_{min} = 2$ | $t_{min} = 3$ | $t_{min} = 4$ | $t_{min} = 5$ |
| Chain | 28.4% | 28.5% | 42.2% | 53.5% | 62.1% | 77.9% |
| Star | 0% | 0% | 0% | 1.3% | 4.7% | 14.2% |
| Loop | 0% | 0% | 0% | 3.6% | 6.2% | 7.6% |

In the smooth phase, traffic efficiency is still well maintained though conflicts occur more frequently. Interestingly, the workload of ATCOs increases with a slower rate as shown in Figure 16 and Table 1. This is one of the most important features of ATCOs' meta-cognition, called "cognition complexity inhibition", which is similar to a proven traffic control strategy "standard flow" (Histon et al., 2002). ATCOs reduce their workload by simplifying the traffic picture and control strategy. As mentioned in Section 4.2.1, in the smooth phase, standard routes are assigned to aircraft to form a stable and familiar traffic structure in order to reduce the controllers' cognition workload. More importantly, in-trailed following of aircraft by speed adjustment makes it much easier to avoid conflicts and reduce communication. Figure 19 (a) shows the dominance of degree 1 in the CSN, suggesting that most conflicts involve only two aircraft. Regarding the degrees of vertices shared by the two networks, their correlations are more significant than the free phase; see Figure 19 (b). The prevailing motifs are chains and loops in the ICN due to the "cognition complexity inhibition" strategy, which is employed in dealing with chain-like or star-like conflicts; see in Table 3.

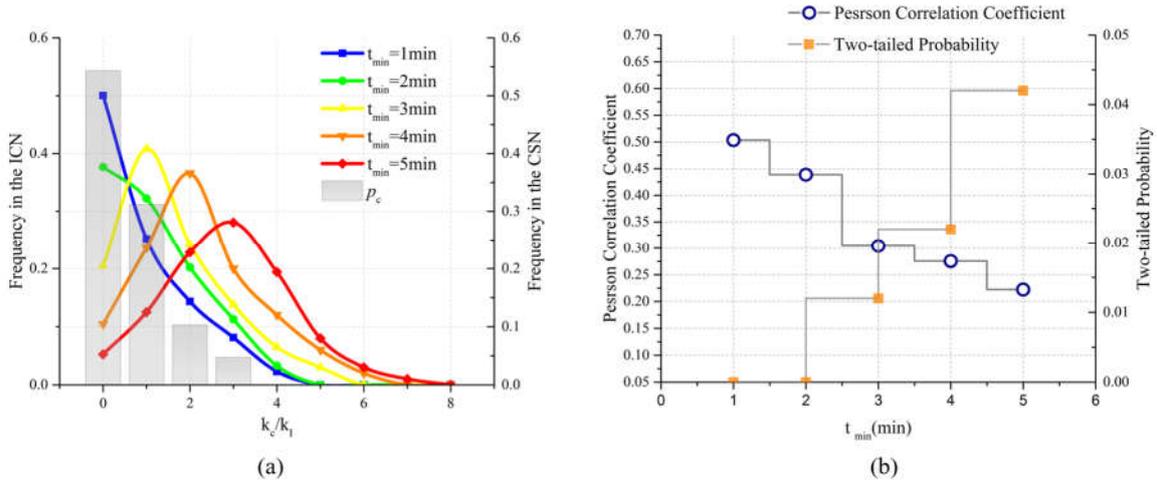

Figure 19. Smooth phase: (a) Degree distributions of CSN and ICN; (b) Degree Correlation for vetices shared by CSN and ICN.

Table 3. Motifs within ICCN in the smooth phase.

| Conflict Situation Network (CSN) | Interventional Communication Network (ICN) | | | | |
|---|---|---|---|---|---|
| | $t_{min} = 1$ | $t_{min} = 2$ | $t_{min} = 3$ | $t_{min} = 4$ | $t_{min} = 5$ |
| Chain | 33.5% | 50.5% | 52.2% | 55.1% | 42.3% | 35% |
| Star | 12.5% | 0% | 0% | 2.5% | 9.8% | 16.9% |
| Loop | 0% | 0% | 9.1% | 12.6% | 36.2% | 42.6% |

In semi-stable phase, the Average Outflow Rate approaches maximum. Sector capacity, defined as the maximum throughput or the threshold of ATCO's workload, is achieved in this phase. As the traffic volume and conflicts significantly increase, the average growth rate of CL is slightly higher than that in the smooth phase due to heading changes adopted to avoid conflicts, which need more interventions than speed adjustment. It implies that driven by metacognition dynamics, the strategy of "cognition complexity inhibition" is inherited from the smooth phase and is applied using "critical points" (e.g. merging points or navigation fixes) along standard routes to ease the traffic picture and workload when issuing radar vectoring commands (Histon et al., 2002). As shown in Figure 20 and Table 4, in the semi-stable phase, as the multi-aircraft conflict grows (see $p_C$), interventional communication is more passive and conflict-oriented, as indicated by the higher degree correlations of shared vertices in each layer. More importantly, facing the increasing complexity of conflict pattern, controllers are still able to solve the problems in an efficient and ordered way (i.e. dominated chain and loop-like motifs of communication network) driven by metacognition dynamics.

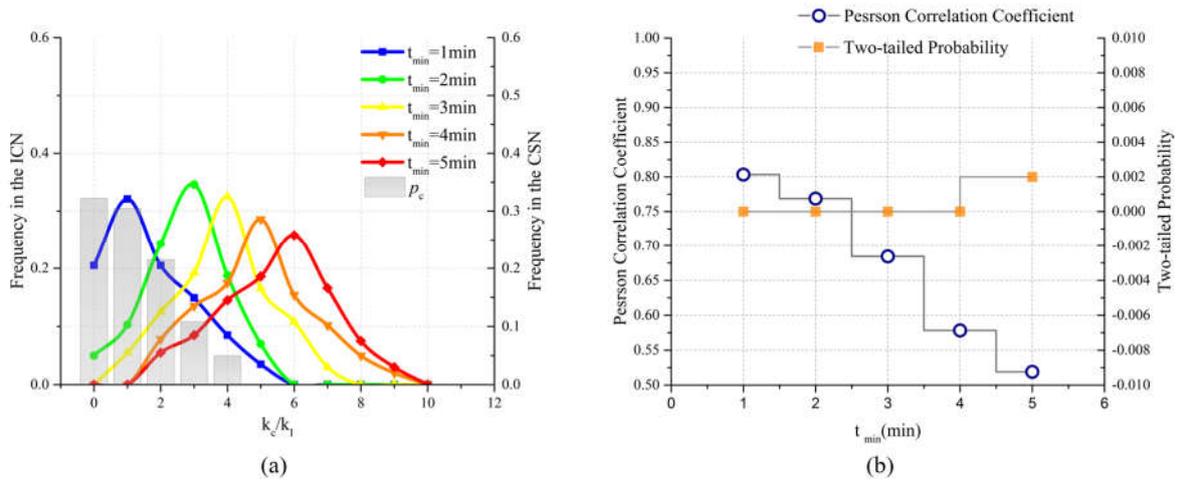

Figure 20. Semi-stable phase: (a) Degree distributions of CSN and ICN; (b) Degree Correlation for vetices shared by CSN and ICN.

Table 4. Motifs within ICCN in the semi-stable phase.

| | Conflict Situation Network (CSN) | Interventional Communication Network (ICN) | | | | |
|---|---|---|---|---|---|---|
| | | $t_{min} = 1$ | $t_{min} = 2$ | $t_{min} = 3$ | $t_{min} = 4$ | $t_{min} = 5$ |
| Chain | 38.4% | 74.5% | 70.2% | 59.1% | 42.3% | 30.2% |
| Star | 20.6% | 0% | 6.5% | 8.3% | 11.5% | 18.5% |
| Loop | 10.7% | 4.8% | 17.1% | 32.6% | 46.2% | 51.3% |

In the congested phase, the Average Outflow Rate drops due to highly complex conflict situations. In order to avoid conflict with tight physical constraints, ATCOs widely adopt heading changes to keep safe separation and maintain flow efficiency. Holding is another standard flow control strategy in congested phases, and reduces the efficiency of air traffic flow due to fixed holding pattern and position. Driven by metacognition dynamics, controllers tend to focus on the safety rather than efficiency in a stressing, or even chaotic, work environment. Figure 21 shows that the degree correlation of shared vertices in the networks are as significant as in the semi-stable phase. Moreover, the increasing complexity of conflict leads to more star motifs in the ICN. Both continuous growth of degree and motifs of ICCN prove the lower efficiency of traffic operation and the consistency of conflict and communication pattern in congested phase.

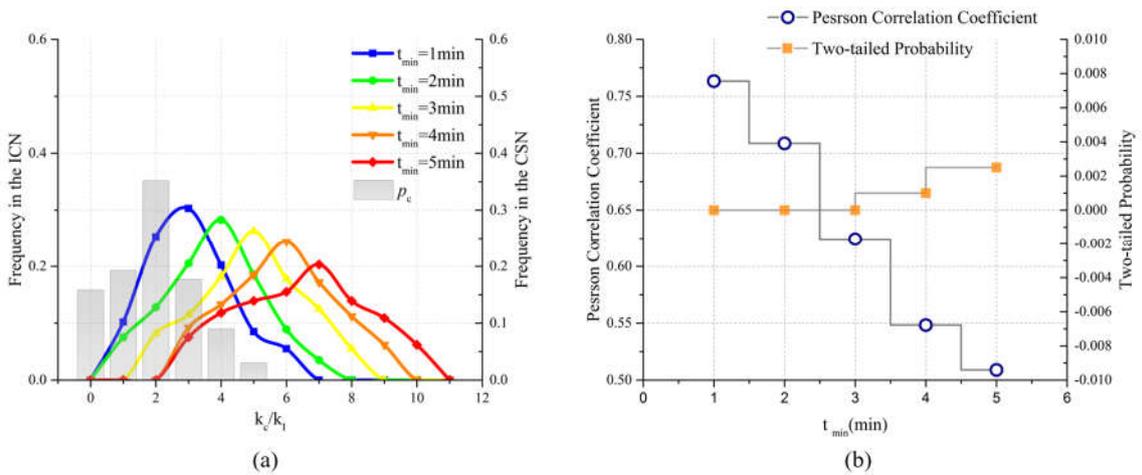

Figure 21. Congested phase: (a) Degree distributions of CSN and ICN; (b) Degree Correlation for vetices shared by CSN and ICN.

Table 5. Motifs within ICCN in the congested phase.

| | Conflict Situation Network (CSN) | Interventional Communication Network (ICN) | | | | |
|---|---|---|---|---|---|---|
| | | $t_{min} = 1$ | $t_{min} = 2$ | $t_{min} = 3$ | $t_{min} = 4$ | $t_{min} = 5$ |
| Chain | 44.4% | 60.5% | 50.2% | 39.1% | 29.4% | 20.6% |
| Star | 25.6% | 15.8% | 22.3% | 28.4% | 34.2% | 38.1% |
| Loop | 14.7% | 23.7% | 27.5% | 32.5% | 36.4% | 41.3% |

In summary, by analyzing the MFD, communication load and network characteristics of ICCN at a sector level, the co-evolution of air traffic situation and ATCOs' performance is quantitatively and qualitatively described. The underlying mechanism of "human-flow" interaction can be reasonably explained as the adaptive cognition management strategy "metacognition dynamics" of air traffic controllers to cope with traffic flow complexity. In the next section, based on the hypotheses and methods proposed in Section 3.3, chaotic properties of the "human-flow" system are explored.

### 4.4. High-level non-linear dynamics of terminal air traffic system

Air traffic system operation is a complex chaotic problem (Shi, 2001). However, non-chaotic feature of air traffic flow was proven in terminal airspace by analyzing the time series of traffic volume (Cong et al., 2014). Nonetheless, intuitively, the dynamic evolution of air traffic flow (in all phases) is the integrated output of adaptive human control activities dealing with increasing traffic conflicts, which exhibits non-linear characteristics such as uncertainty, burstiness and diffusivity. Besides, potential conflicts can be regarded not only as the dynamics of air traffic demand but also the system emergence triggered by "human-flow" and "human-human" interactions during multi-sectors operation (e.g. local conflict resolution in one sector or one route will lead secondary conflict in other area). With these in mind, two hypotheses are made as follows.

***Hypothesis 1***: the system under investigation is chaotic, and the chaotic phenomenon can be observed in both terminal and sector levels.

***Hypothesis 2***: Chaos is highly related to the phase of air traffic flow.

*4.4.1. Chaotic analysis of air traffic flow at the terminal-level*

We choose traffic on 11/09/2012 as a sample to present the results of chaotic analysis according to the method in Section 3.3. Figure 22 shows the 24-hour time series of potential conflicts with $\tau = 5min$.

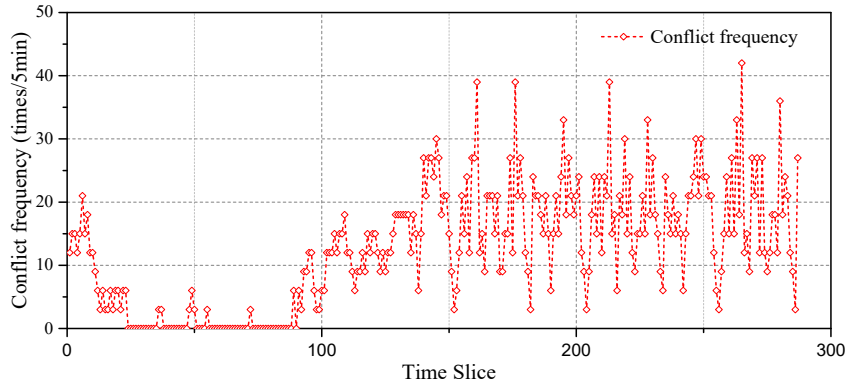

Figure 22. Time series of potential conflict.

By adopting autocorrelative function, the delay time $\lambda_{\mathcal{H}} = 23$ is obtained when function value reaches the minimum for the first time as shown in Figure 23(a). Meanwhile, false nearest neighbor algorithm is used to calculate the embedded dimension $\mathcal{M}_{\mathcal{H}}$. When the decreasing rate of the proportion of false nearest neighbor points is less than 0.001, the attractors are regarded as unfolded (Maus et al., 2011, Farmer et al., 1983). Figure 23 (b) shows the evolution of proportion of false nearest neighbor points, the embedded dimension of the time series is $\mathcal{M}_{\mathcal{H}} = 7$. Then, the largest Lyapunov exponent is calculated in order to assess the system's sensitivity to initial conditions. Chaos features is identified as the value of the largest Lyapunov exponent is 0.00193 (>0) as shown in Figure 23 (c). It is noted that the same conclusions of chaotic system are drawn by analyzing traffic data on 10/03/2012 and 15/12/2012 with the largest Lyapunov exponent is 0.00174 and 0.00202 respectively.

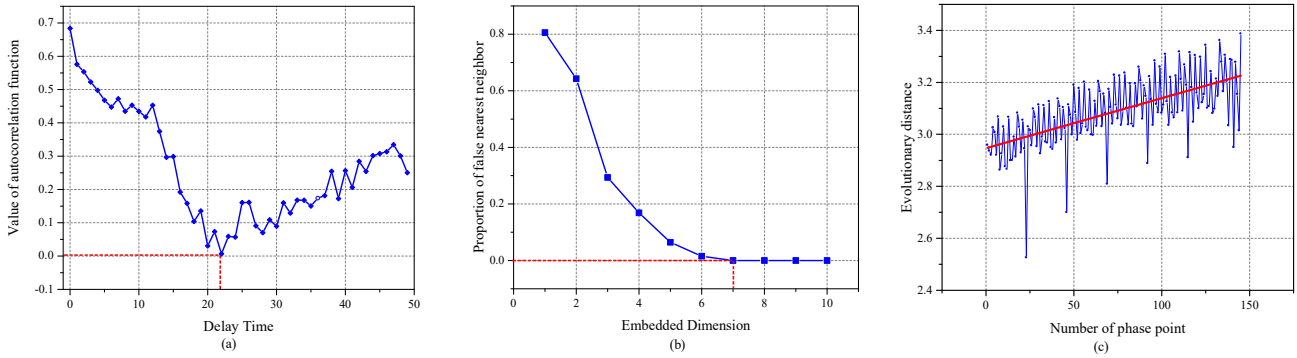

Figure 23. Result of chaotic analysis. (a) Delay time calculation. (b) Embedded dimension measurement.(c) The largest Lyapunov exponent (calculated as the slope of the best linear fit)

We have initially demonstrated that air traffic flow at terminal level is chaotic represented by system variable of potential conflicts. To further reveal the chaotic evolution with traffic volume, we calculate the largest Lyapunov exponent every 4 hours in the three days, and show the results in Figure 24. It is shown that chaos in the system is induced by high traffic and potential conflicts, which created more random elements under multi-agent interactions.

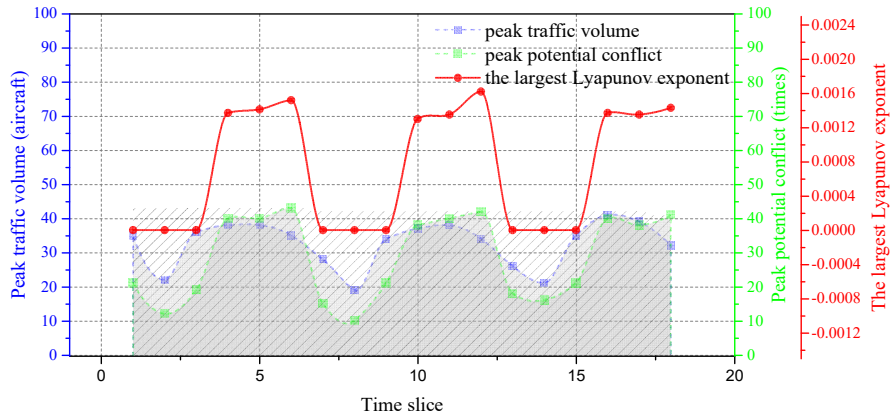

Figure 24. Chaotic evolution with traffic volume and potential conflict

### 4.4.2. Chaotic analysis of "human-flow" system at the sector level

Since air traffic system can be simply divided into "flow system" and "human-system", to further understand the chaotic features of this artificial system under "human-flow" interactions, additional data of ATCO's behavior is studied by adopting method in Section 3.3 together with time series of potential conflicts in specific sectors. Inspired by previous research on ATCOs' communication dynamics (Wang et al., 2013) and chaos analysis of vehicle headway evolution at a fixed observing spot (Zhang et al., 2009), silence period of radio channel (a.k.a. *communication interval*), which characterizes temporal dynamics of controllers (Wang et al., 2013), is used as a representative of "human system". Here, both communication types of intervention and nonintervention are included. For simplicity, the series of silence period and potential conflict of Sector 2 on 11/09/2012 are shown in Figure 25 (a)-(b). Result shows that the largest Lyapunov exponents of silence period series and potential conflict time series of Sector 2 are 0.00131 and 0.00173 respectively as shown in Figure 25 (c)-(d).

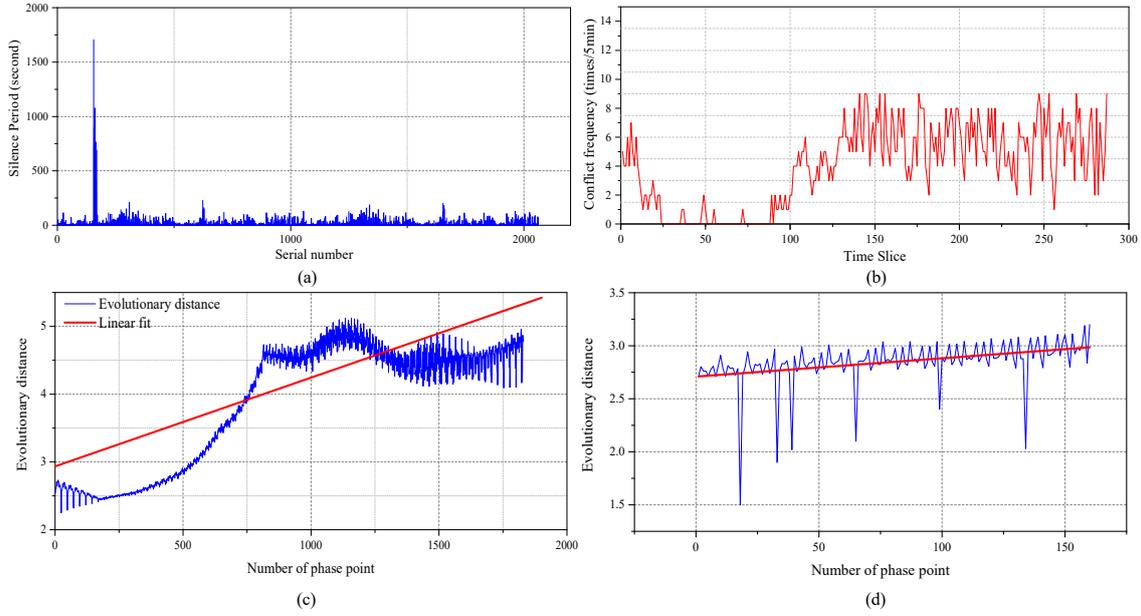

Figure 25. Selected input and output of chaotic analysis in Sector 2. (a) Silence period series; (b) Conflict time series; (c) The largest Lyapunov exponent of silence period series; (d) The largest Lyapunov exponent of potential conflict series.

The same method is applied to identify the chaos features in the 4 other sectors respectively. Table 6 shows the result of delay time, embedded dimension and the largest Lyapunov exponent of all 5 sectors on the day of 11/09/2012. Interestingly, chaos in "human system" and "flow-system" emerges asynchronously: flow chaos emerges in sectors 1, 2, 3 and 5 where semi-stable or/and congested phases are observed, while human chaos is absent in sector 3, where there is no congested phase. Based on the discussions on underlying mechanism of "human-flow" interactions in Section 4.4, conclusions of asynchronism can be simply supported by ATCOs metacognition of avoiding chaos as the increase of traffic. It is noted that the same conclusion of asynchronous chaos of "human system" and "flow-system" are drawn by analyzing the data on the other two days.

Table 6. Chaotic results of "human-flow" system in airspace sectors

|  | Sector 1 | | Sector 2 | | Sector 3 | | Sector 4 | | Sector 5 | |
|---|---|---|---|---|---|---|---|---|---|---|
|  | Flow | ATCO | Flow | ATCO | Flow | ATCO | Flow | ATCO | Flow | ATCO |
| Delay time | 21 | 20 | 22 | 24 | 17 | 9 | 8 | 5 | 23 | 22 |
| Embedded dimension | 7 | 9 | 8 | 10 | 6 | 6 | 2 | 3 | 9 | 11 |
| The largest Lyapunov exponent | 0.00155 | 0.00129 | 0.00173 | 0.00131 | 0.00101 | 0 | 0 | 0 | 0.00149 | 0.00127 |

By analyzing the chaotic dynamics at terminal and sector levels, we conclude that chaos exists in terminal airspace, and emerges when traffic becomes (nearly) saturated; it is intrinsic to the dynamics of the co-evolution of "flow system" and "human-system". Due to the highly congested airspace systems in China, European and the U.S., it is expected that chaotic dynamics may be prevalent and provide new tools and insights for the prediction and control of air traffic system evolution.

## 5. Conclusions

The study of complex air traffic dynamics is essential for understanding the nature of air traffic system and uncovering technical potentials for advancing air traffic management. Among a variety of analytical approaches, network analysis is an effective and intuitive way to understand multi-agent behavior at varying granularities. In order to characterize air traffic physics and interpret underlying mechanism of "ATCOs-flow" interactions, a comprehensive multi-layer network is constructed to integrate and present elements like airspace, aircraft trajectory, flow transmissions, potential conflicts and ATCOs' interventions activities. To further represent the dynamics of "human-flow" interaction at different granularities, as well as system nonlinearity, a series of analytical metrics of air traffic flow and human controller are introduced (e.g. network flow variables, ATCOs' cognitive complexity, and chaos).

Systematic analyses are conducted based on the proposed network as well as analytical metrics using a case study of Guangzhou terminal airspace; the main subjects of investigation include

1. network flow dynamics: FD, MFD and four traffic phases: free, smooth, semi-stable, and congested
2. underlying mechanism of "ATCOs-flow" interactions: Meta-cognition strategies of controllers are interpreted as the mechanism of "ATCOs-flow" performance co-evolution including "pre-activation", "inhibition" and "stress"
3. Chaos identification: at the system level, chaos is identified in both flow and human dynamics, when air traffic is characterized as semi-stable or congested.

This study represents an empirical exploration of the Guangzhou terminal airspace by employing a generic methodology. As a future study, it is essential to test the generality of our findings and identify critical features that are responsible for qualitatively different system behaviors. Real-time human-in-the-loop simulation is considered an effective way to supplement addition of ATCO-related data in various scenarios.

# Appendix A. RAN-FTN mapping algorithm

We detail the algorithm that maps the operational flight paths to the RAN, as summarized in Table 7 and illustrated in Figure 26.

Table 7. Algorithm of generating cross-layer links between RAN and FTN

| Algorithm 1. Connectivity Generation between RAN and FTN vertices |
| --- |
| **Step1:** Extract sets of standard flight paths and operational flight paths, denoted as $\mathcal{P}_{RAN}$ and $\mathcal{P}_{OFP}$, respectively. |
| **Step2:** For each OFP, simplify its structure by identifying and grouping its Turning Points (TPs) according to Section A.1. |
| **Step3:** Find the unique matches between each simplified OFP and SFP by following Section A.2. |

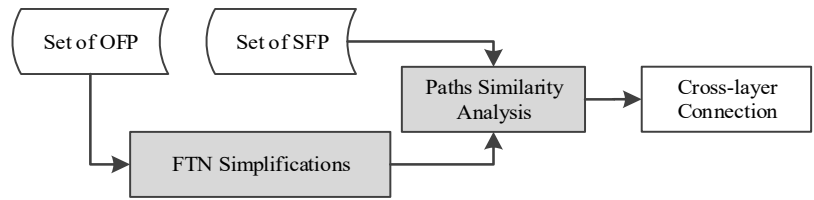

Figure 26. Schema of generating edges between FTN and RAN Vertices

## A.1. Simplification of operational flight path

An operational flight path, which consists of numerous points recorded by the radar, may be approximated by a sequence of line segments mixed with turning maneuvers (see Figure 27). Such a point of view could tremendously simplify the analysis and computation of network dynamics, and is the main motivation of this section.

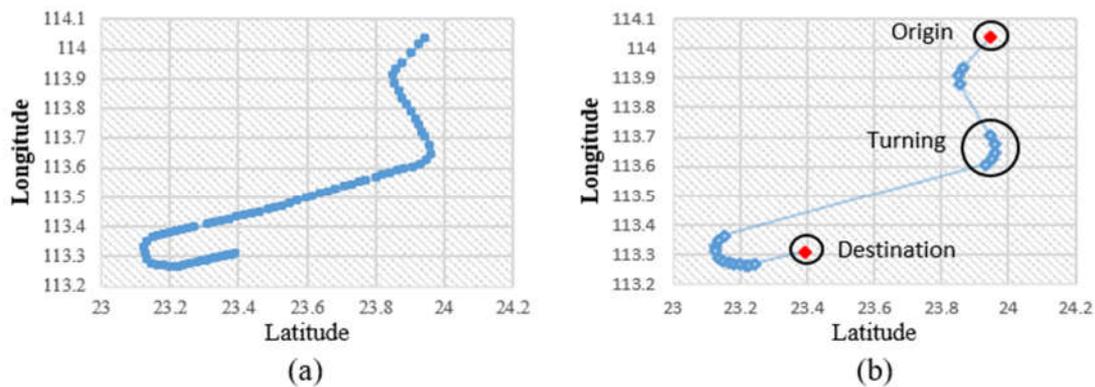

Figure 27. OFP simplification: (a) Original OFP; (b) Simplified OFP consisting of OD points and turning groups

We consider a path in the FTN expressed as a sequence of points $p^l = (x^l, y^l) \in \mathbb{R}^2$, $l = 1, ..., L$; see Figure 27(a). A turning is detected between two consecutive segments $\overrightarrow{p^{l-1}p^l}$ and $\overrightarrow{p^l p^{l+1}}$ if the angle formed by them is above a threshold (Gariel et al., 2011). Mathematically, the turning is confirmed if

$$\phi^l \doteq \left| \arctan\left[\frac{y^{l+1} - y^l}{x^{l+1} - x^l}\right] - \arctan\left[\frac{y^l - y^{l-1}}{x^l - x^{l-1}}\right] \right| > \phi_0 \tag{9}$$

Here, the threshold $\phi_0 = 2.5°$ is experimentally set following Gariel et al. (2011). When (9) holds, we call $p^l$ a turning point. Depending on the time/space headway of the points that constitute a path, an aircraft turning maneuver may contain several consecutive turning points. Accordingly, we define a turning group to be the maximum set that contains a sequence of consecutive turning points. The resulting OFP can be expressed by a sequence of turning groups and straight flight segments, which we call *simplified* OFP; see Figure 27(b). The network consisting of simplified OFP is called simplified FTN.

## A.2. Path Similarity Analysis

In this section, we propose an efficient method based on weighted line segment distance to find the best match of vertices in the RAN and simplified FTN.

### A.2.1. Similarity measurement for line segments

The distance between two line segments in a two-dimensional plane can be assessed by three quantities: perpendicular distance, parallel distance, and angle distance, as given by formulae (10), (11), and (12), respectively. These three components are adapted from similarity measures used in pattern recognition (Chen et al, 2003). We define these distances based on Figure 28, without loss of generality.

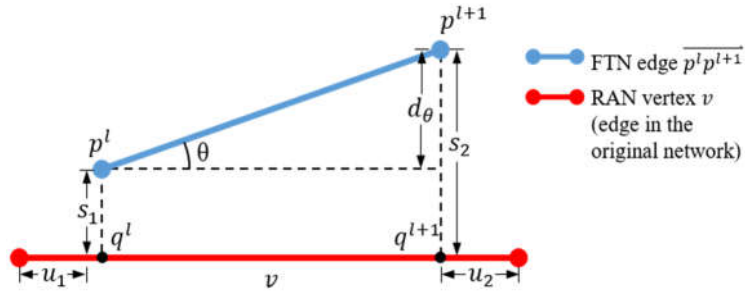

Figure 28. Different distances of two line segments

Let $s_1$ and $s_2$ be the distances of $p^l$ and $p^{l+1}$ from $v$, respectively. The *perpendicular distance* between the segments $\overrightarrow{p^l p^{l+1}}$ and $v$ is defined as

$$d_p\left(\overrightarrow{p^l p^{l+1}}, v\right) \doteq \frac{s_1^2 + s_2^2}{s_1 + s_2} \tag{10}$$

The *parallel distance* is defined as:

$$d_l\left(\overrightarrow{p^l p^{l+1}}, v\right) = \min(u_1, u_2) \tag{11}$$

The *angle distance* is defined as:

$$d_\theta\left(\overrightarrow{p^l p^{l+1}}, v\right) = \begin{cases} \|p^{l+1} - p^l\| \times \sin\theta & 0 \leq \theta < \pi/2 \\ \|p^{l+1} - p^l\| & \pi/2 \leq \theta \leq \pi \end{cases} \tag{12}$$

Finally, the line segment distance is a convex combination of the distances above.

$$d\left(\overrightarrow{p^l p^{l+1}}, v\right) = \alpha d_p\left(\overrightarrow{p^l p^{l+1}}, v\right) + \beta d_l\left(\overrightarrow{p^l p^{l+1}}, v\right) + \gamma d_\theta\left(\overrightarrow{p^l p^{l+1}}, v\right) \tag{13}$$

where $\alpha, \beta, \gamma \geq 0$, $\alpha + \beta + \gamma = 1$. Then, the similarity of the two segments is quantified as follows with a given threshold $\delta$ (experimentally set as 10km in this paper):

$$sim\left(\overrightarrow{p^l p^{l+1}}, v\right) = \begin{cases} 0, & \text{if } d\left(\overrightarrow{p^l p^{l+1}}, v\right) > \delta \\ 1 - \frac{d\left(\overrightarrow{p^l p^{l+1}}, v\right)}{\delta}, & \text{otherwise} \end{cases} \tag{14}$$

*A.2.2. Similarity of paths in the RAN and FTN*

After defining the similarity between pairs of line segments, we move on to the path similarity measure. We introduce a recursively defined function to find the longest common subsequence (LCS) of each turning trajectory, which is a modification of the existing methods (Hermes et al., 2009; Vlachos et al., 2002). A simplified (partial) OFP in the FTN is expressed as a sequence of links, denoted by $P^k = \{a^0, a^1, ..., a^{k-1}\}$, $k = 0,1,...$. Given a path in the RAN $R^h = \{\aleph^0, \aleph^1, ..., \aleph^h\}$, $h = 0,1,...$. We defined the similarity of LCS as follows.

$$simLCS(P^k, R^h) = \begin{cases} 0, & \text{if } k = 0 \text{ or } h = 0 \\ \max\{simLCS(P^{k-1}, R^{h-1}) + sim(a^{k-1}, \aleph^h), \\ simLCS(P^{k-1}, R^h), simLCS(P^k, R^{h-1})\} & \text{otherwise} \end{cases} \quad (15)$$

This function returns the similarity measure of best alignment between $P^k$ and $R^h$. Then, the path in the RAN that optimally matches the OFP $P^k$ is selected.

## Acknowledgments

This research was supported by the National Natural Science Foundation of China (Grant No. F030209), and the Chinese Scholarship Council (CSC).